\documentclass[a4paper, notoc]{JHEP3}
\usepackage{amsfonts}
\usepackage{amssymb}
\usepackage{comment}
\usepackage{epsfig}
\usepackage{marginnote}
\usepackage{graphicx}
\usepackage{amsmath}

\newcommand{\be}{\begin{equation}}
\newcommand{\ee}{\end{equation}}
\newcommand{\bea}{\begin{eqnarray}}
\newcommand{\eea}{\end{eqnarray}}
\newcommand{\mbb}{\mathbb}

\newcommand{\mc}{\mathcal}

\newcommand{\beqa}{\begin{eqnarray}}
\newcommand{\eeqa}{\end{eqnarray}}

\newcommand{\vo}{{\cal{V}}}
\newcommand{\I}{{\rm i}}

\newcommand{\sphantom}{&}

\def\pref#1{(\ref{#1})}

\title{Toric K3-Fibred Calabi-Yau Manifolds with del Pezzo Divisors for String Compactifications}

\author{Michele Cicoli${}^1$, Maximilian Kreuzer, Christoph Mayrhofer${}^2$ \\

$^1$ Deutsches Elektronen-Synchrotron DESY, Notkestrasse 85, 22607 Hamburg, Germany.\\
Email: \email{michele.cicoli@desy.de} \\
$^2$ Institut f\"ur Theoretische Physik, Universit\"at Heidelberg, Heidelberg, Germany.\\
Email: \email{c.mayrhofer@thphys.uni-heidelberg.de}}

\abstract{We analyse several explicit toric examples of compact K3-fibred Calabi-Yau three-folds. These manifolds can be used for the study of string dualities and are crucial ingredients
for the construction of LARGE Volume type IIB vacua with promising applications to cosmology and particle phenomenology.
In order to build a phenomenologically viable model, on top of the two moduli corresponding to the base and the K3 fibre,
we demand also the existence of two additional rigid divisors:
the first supporting the non-perturbative effects needed to achieve moduli stabilisation, and
the second allowing the presence of chiral matter on wrapped D-branes.
We clarify the topology of these rigid divisors
by discussing the interplay between a diagonal structure of the Calabi-Yau volume and D-terms.
Del Pezzo divisors appearing in the volume form in a completely
diagonal way are natural candidates for supporting non-perturbative effects
and for quiver constructions, while
`non-diagonal' del Pezzo and rigid but not del Pezzo divisors
are particularly interesting for model building in the geometric regime.
Searching through the existing list of four dimensional reflexive lattice polytopes,
we find 158 examples admitting a Calabi-Yau hypersurface
with a K3 fibration and four K\"ahler moduli where at least one of the toric divisors
is a `diagonal' del Pezzo. We work out explicitly the topological details of a few examples
showing how, in the case of simplicial polytopes, all the del Pezzo divisors are `diagonal', while `non-diagonal' ones appear only in the case of non-simplicial polytopes.
A companion paper will use these results in the study of moduli stabilisation for globally
consistent explicit Calabi-Yau compactifications with the local presence of chirality. \\
\begin{center}
{\em Dedicated to the memory of Maximilian Kreuzer}
\end{center}}

\preprint{DESY 11-103}

\begin{document}

\tableofcontents

\bigskip

\section{Introduction}

K3 fibrations play a crucial r\^ole in the study of string compactifications.
Historically they have been first used to study mirror symmetry for two-parameter Calabi-Yau manifolds
built as hypersurfaces embedded in complex weighted projective spaces \cite{Candelas:1993dm}.

In \cite{Klemm:1995tj} it has then been realised that K3 fibrations are a fundamental ingredient to understand strong-weak
coupling dualities between 4D $\mc{N}=2$ supersymmetric theories
which are the low energy result of string compactifications \cite{Kachru:1995wm,Sen:1995ff}.
Subsequently, it has been shown that if some strongly coupled $\mc{N}=2$ heterotic vacua admit a dual
description in terms of weakly coupled type IIA string theory
compactified on a Calabi-Yau three-fold, the compactification manifold has to be a K3 fibration \cite{Aspinwall:1995vk}
(see also \cite{Aldazabal:1995yw,Hunt:1995sy,Aspinwall:1996mw,Gomez:1996xi} for other studies of string dualities based on K3 fibrations).

The original work of \cite{Klemm:1995tj} provided a short list of K3-fibred Calabi-Yau three-folds,
consisting of 31 hypersurfaces embedded in weighted $\mbb{P}^4$ and 25 complete intersections in weighted $\mbb{P}^5$.
These results were generalised in \cite{Hosono} where the authors found 628 examples identifying all the K3 fibrations which can be
built as transverse polynomials in weighted $\mbb{P}^4$. The powerful methods of toric geometry
allowed the authors of \cite{Avram:1996pj} to provide many more explicit constructions of K3-fibred Calabi-Yau
three-folds. In fact, they found 124,701 examples of K3 fibrations realised as hypersurfaces embedded in toric varieties.

More recently, K3-fibred Calabi-Yau three-folds received new attention in the context of type IIB orientifold compactifications.
Here all the closed string moduli, but those parameterising the K\"ahler deformations of the Calabi-Yau metric,
can be stabilised by turning on background fluxes \cite{Giddings:2001yu}.

Very interesting vacua among this kind of string constructions are the so-called LARGE Volume Scenarios (LVS),
which are characterised by the presence of a non-supersymmetric minimum at exponentially large volume
of the compactification manifold \cite{Balasubramanian:2005zx,Cicoli:2008va}.
The simplest LVS, originally proposed in \cite{Balasubramanian:2005zx}, is based on
type IIB compactified on an orientifold of the Calabi-Yau $\mbb{P}^4_{[1,1,1,6,9]}(18)$.
This Calabi-Yau admits only two K\"ahler moduli which can be stabilised by the interplay of
non-perturbative and $\alpha'$ corrections without the need to fine-tune the background fluxes.
The internal volume $\vo = \tau_b^{3/2} -\tau_s^{3/2}$
is controlled by one exponentially large 4-cycle $\tau_b\sim e^{c/g_s}\gg 1$ (for $c\sim\mc{O}(1)$ and $g_s< 1$ in order to
trust perturbation theory). The
second K\"ahler modulus $\tau_s\sim 1/g_s>1$ controls the volume of a small blow-up mode.
The fact that the Calabi-Yau volume is exponentially large
guarantees that the approximations in the effective field theory
are under good control and allows to generate several interesting
phenomenological hierarchies.

This simple construction can be generalised adding more blow-up modes which can all be fixed at relatively small values by
non-perturbative effects \cite{Conlon:2005ki}. The Calabi-Yau acquires a distinctive `Swiss-cheese' shape
where the overall volume, $\vo = \tau_b^{3/2}  -\sum_i\tau_{s,i}^{3/2}$, is still controlled by just one 4-cycle $\tau_b$
which can be arbitrarily large while the various small four-cycles $\tau_i$ control the
size of the `holes' of the Swiss-cheese. Examples with $h^{1,1}=3$ having this form are the Fano three-fold
$\mc{F}_{11}$ \cite{Denef:2004dm}, the degree 15 hypersurface embedded in
$\mbb{P}^4_{[1,3,3,3,5]}$ and the degree 30 hypersurface in
$\mbb{P}^4_{[1,1,3,10,15]}$ \cite{Blumenhagen:2007sm} while several examples with $h^{1,1}=4$
have been provided in \cite{Collinucci:2008sq}.

The first examples of LVS with a topological structure more
complicated than the Swiss-cheese one, were discovered in
\cite{Cicoli:2008va}. The authors assumed the existence of a K3-fibred Calabi-Yau with
$h^{1,1} = 3$, obtained by adding an \textit{ad hoc} blow-up mode to the
geometry $\mbb{P}^4_{[1,1,2,2,6]}(12)$ studied in \cite{Candelas:1993dm}.
Now the Calabi-Yau volume is not controlled anymore by just one large four-cycle,
but by two four-cycles: the K3 fibre, $\tau_1$ and another four-cycle $\tau_2$ which
contains the $\mbb{P}^1$ base $t_1$ of the fibration. Hence it takes the form
$\vo = \sqrt{\tau_1}\tau_2 - \tau_s^{3/2}
 = t_1\tau_1-\tau_s^{3/2}$
with $\tau_2= t_1 \sqrt{\tau_1}$
(the generalisation to models with many blow-up modes is straightforward).

The fact that the internal volume is controlled by two four-cycles turns out
to be very promising for both cosmological and phenomenological applications
for the following reasons:
\begin{itemize}
\item Given that $\alpha'$
and non-perturbative effects fix the overall volume exponentially large,
only the combination $\sqrt{\tau_1}\tau_2\sim e^{c/g_s}\gg 1$ gets
stabilised leaving a flat direction (the small blow-up mode is still fixed at
relatively small values $\tau_s\sim 1/g_s>1$).
Other stabilising effects come from string loop corrections to the K\"ahler
potential but it has been shown that for the original $\mbb{P}^4_{[1,1,1,6,9]}(18)$ model
they turn out to be subleading with respect to the other corrections \cite{Berg:2007wt}.
The authors of \cite{Cicoli:2007xp} then showed that
$g_s$ corrections turn out generically to be subdominant with respect to $\alpha'$
and non-perturbative effects. This is due to a leading order cancellation
of the $g_s$  contribution to the scalar potential, called `extended no-scale structure'.

Therefore, in the K3-fibred case, there is a direction in the $(\tau_1,\tau_2)$-plane
orthogonal to the overall volume which is naturally flatter than all the other ones. This remaining flat direction can be lifted only when subleading string loops are taken into account.
In fact, these corrections give rise to a minimum located at $\tau_1 \sim g_s^{4/3}\,\vo^{2/3}$ \cite{Cicoli:2008va}.

Due to the shallowness of this direction, this scalar field is a natural candidate
for cosmological applications. In \cite{Cicoli:2008gp} it has been used as an inflaton field,
in a large-field inflationary model which predicts
observably large gravitational waves and
where the $\eta$-problem can be solved without requiring fine-tuning.
Subsequently, this direction has also been exploited as a curvaton field,
leading to an interesting model which predicts detectable non-Gaussianities of local type \cite{Burgess:2010bz}.

Other interesting cosmological applications of LARGE Volume K3-fibred
compactifications involve the study of reheating \cite{Cicoli:2010ha,Cicoli:2010yj}
and finite-temperature effects \cite{Anguelova:2009ht}.

\item The \textit{base $\times$ fibre} structure of the volume gives the possibility
to take the anisotropic limit $t_1 \gg \sqrt{\tau_1}$ (or $\tau_2 \gg \tau_1$),
corresponding to interesting geometries having the two dimensions of the base,
spanned by the cycle $t_1$, larger than the other four of the K3
fibre, spanned by $\tau_1$.

These anisotropic compactifications turn out to be very promising
to make contact with current experiments
from the study of the phenomenological properties of hidden Abelian gauge bosons
with a kinetic mixing with the ordinary photon.
In fact, they lead naturally to dark forces for an intermediate string scale,
or even to a hidden CMB allowing some fine-tuning of the underlying parameters
for the extreme case of TeV-scale strings \cite{Cicoli:2011yh}.

Moreover, it has been recently shown that the moduli can indeed be stabilised in
this very anisotropic region of the K\"ahler cone \cite{Cicoli:2011yy}. This
leads to the first stringy derivation of popular models
with effectively two large extra dimensions of micron-size and
fundamental gravity scale around a few TeV \cite{ArkaniHamed:1998rs,Antoniadis:1998ig,Aghababaie:2003wz,Aghababaie:2003ar}.
The use of K3-fibred compactifications is needed not only to
obtain an anisotropic shape of the Calabi-Yau but also
to generate the poly-instantons effects \cite{Blumenhagen:2008ji} which are
crucial to fix the K3 divisor at relatively small values $\tau_1\sim 1/g_s>1$
like a blow-up mode of the previous examples.
\end{itemize}
Despite all these phenomenological successes, these models still lack a rigorous mathematical
foundation, in the sense that no explicit Calabi-Yau examples of K3 fibrations
with additional blow-up modes are known in the literature yet. The goal of this paper
is exactly to fill in this gap by using the powerful methods of toric geometry.

We shall search for K3-fibred Calabi-Yau three-folds which satisfy the two topological
conditions for the existence of a LVS minimum derived in \cite{Cicoli:2008va}:
\begin{enumerate}
\item{} The Euler number of the Calabi Yau manifold must be
negative, or more precisely: $h^{1,2} > h^{1,1} > 1$;

\item{} The Calabi-Yau manifold must have at least one `diagonal' blow-up mode.
\end{enumerate}
We shall discuss the second condition and understand its geometrical meaning
more in depth. In fact, in section \ref{sec:dia-divisor} we shall notice that there can be three types of `blow-up'
divisors:
\begin{itemize}
\item \emph{`Diagonal' del Pezzo}: these four-cycles are truly local effects, in the sense that
it can always be found a basis where this divisor $D_{{\rm dP}}$ has only its triple self intersection number
non-vanishing, or equivalently where $\sqrt{{\rm Vol}\left(D_{{\rm dP}}\right)}\propto {\rm Vol}\left(D_{{\rm dP}}\cap D_{{\rm dP}}\right)$.
This conditions implies that this del Pezzo four-cycle will appear in the volume form in a completely
diagonal manner. It can also be shown to correspond to the condition on the inverse K\"ahler metric
$K^{-1}_{{\rm dP},{\rm dP}} = \vo\sqrt{\tau_{{\rm dP}}}$, where $\tau_{{\rm dP}}={\rm Vol}\left(D_{{\rm dP}}\right)$. This condition has been used in
\cite{Cicoli:2008va} to infer that the divisor $D_{{\rm dP}}$ has to come from a resolution of a point like singularity.

Besides being the natural candidate for supporting the non-perturbative effects of the LVS,
this kind of divisor is very useful also for chiral model building. In fact, in section~\ref{sec:d-terms}
we shall show that the constraint coming from the
vanishing $D$-term condition, in the absence of singlets which can acquire non-vanishing
vacuum expectation values, always forces this four-cycles to shrink to zero size. This
naturally leads to quiver GUT-like constructions \cite{Conlon:2008wa,Blumenhagen:2009gk}.

\item \emph{`Non-diagonal' del Pezzo}: these four-cycles are characterised by the fact that it is
impossible to find a basis where they do not intersect any other divisor. Given that they are not
genuinely local effects, they will not appear in the Calabi-Yau volume in a completely diagonal way,
corresponding to a form of the relative elements of the inverse K\"ahler metric which
renders these four-cycles inappropriate for supporting the non-perturbative effects needed
for the realisation of the LVS.

On the contrary, these divisors turn out to be the best candidates to
support GUT constructions with magnetised branes wrapping cycles stabilised in the geometric
regime (see \cite{Blumenhagen:2008zz} for the type IIB case and \cite{Beasley:2008dc,Beasley:2008kw}
for the F-theory case). For rigid divisors no unwanted matter in the adjoint representation
gets generated. Furthermore, in this `non-diagonal' case,
the vanishing $D$-term condition may not force the shrinking of any cycle.
However, the final answer to this question is very model-dependent given that
some examples of this kind are known where the $D$-terms still lead the system
to the boundary of the K\"ahler cone \cite{Collinucci:2008sq}.

\item \emph{Rigid but not del Pezzo}: there can finally be other rigid four-cycle with $h^{1,0}=h^{2,0}=0$
which are not del Pezzo divisors, meaning that they are not contractible to a point. An intuitive
way to visualise these four-cycles, is to think about them as blow-ups of curve.
Similarly to the `non-diagonal' del Pezzo divisors, they also cannot be used
to support the non-perturbative effects in the LVS, but are natural candidates
for GUT model building in the geometric regime.
\end{itemize}
With this classification of blow-up cycles at hands, we performed a search
through the existing list of four dimensional reflexive lattice polytopes~\cite{Kreuzer:2000xy,cydata}
for K3-fibred Calabi-Yau three-folds with $h^{1,1}=4$ where at least one
of the toric divisors is a `diagonal' del Pezzo divisor.
As we have already pointed out, this condition is needed in order to obtain the LVS.
On top of the presence of the two moduli corresponding to the base and
the K3 fibre and the four-cycle for the non-perturbative effect, we demand the presence of a second rigid divisor for chiral model building.

The final result of our scan, summarised in section~\ref{sec:k3search}, consists of 158 examples of lattice polytopes.
In the case of simplicial polytopes, all the del Pezzo
four-cycles of each Calabi-Yau example are `diagonal' divisors, as can be seen in the illustrative exemplar worked out in detail in section~\ref{sec:simp-polytopes}.

In the the $h^{1,1}=4$ case, `non-diagonal' del Pezzo divisors appear only in the case of non-simplicial polytopes. We discuss them in section~\ref{sec:non-simp-polytopes}. There, we present two examples of K3 fibrations,
the first with two del Pezzo four-cycles whereas the second with just a single del Pezzo divisor. Two more explicit examples are described in appendix~\ref{appA}.
Finally in appendix~\ref{sec:the-list} we present the full list of all the 158 examples of lattice polytopes
which have a Calabi-Yau hypersurface with $h^{1,1}=4$ that admits a K3 fibration structure and at least one `diagonal' del Pezzo.

\section{Diagonal divisors}
\label{sec:dia-divisor}

In this section we shall present a procedure to single out the divisors which enter the
volume form in a completely diagonal way~\cite{diss}, and hence the natural candidates to
support the non-perturbative effects of the LVS.

\subsection{K\"ahler moduli of Calabi-Yau compactifications}

Before describing the details of the
diagonalisation procedure, let us set our notations and conventions focusing on the
case of type IIB flux compactifications on an orientifold of a Calabi-Yau three-fold $Y_3$,
which preserves $\mc{N} = 1$ supersymmetry in 4D \cite{Giddings:2001yu,Douglas:2006es,Denef:2007pq}.

Considering orientifold
projections such that $h^{1,1}_- = 0 \Rightarrow h^{1,1}_+ =
h^{1,1}$, the K\"ahler form, $J$, can be expanded in a basis
$\{ \hat{D}_i \}_{i=1}^{h^{1,1}}$ of $H^{1,1}(Y_3,\mbb{Z})$
 as:
\be
J = t^i \hat{D}_i.
\ee
Each two-form $\hat{D}_i$ is Poincar\'e dual to the divisor $D_i$ on $\mc X$, i.e.:
\be
\int_{\mc{X}} \omega \wedge \hat{D}_i := \int_{D_i} \omega  \quad
\forall\,\omega \in H^{\textmd{dim}(\mc X)-2}(\mc X, \mbb{Z})\quad \textmd{and} \quad {D}_i\in H_{\textmd{dim}(\mc X)-2}(\mc X, \mbb{Z})\,,
\ee
where $\mc X$ can be a Calabi-Yau hypersurface $Y_3$ or, its toric ambient space $X_4$.
We shall keep the `hat' in this section but drop it in the following, given that it should always be clear from the context whether it is a divisor or its Poincar\'e dual two-form.

The volume of the internal manifold takes the form:
\be
  \vo = \frac{1}{3!}\int_{Y_3} J\wedge J\wedge
  J = \frac{1}{3!} \, k_{ijk} t^i t^j t^k. \label{IlVolume}
\ee
Here $k_{ijk}$ are the triple intersection numbers of
$Y_3$ given by:
\be
k_{ijk}=\int_{Y_3}\hat{D}_i\wedge \hat{D}_j\wedge \hat{D}_k={D}_i\cdot {D}_j\cdot {D}_k\,,
\ee
while the $t^i\in \mbb{R}$ parameterise 2-cycle volumes. Note that the $t^i$ are not restricted to the positive real numbers since the $\hat D_i$ are not in general the generators of the K\"ahler cone.

The quantities $t^i$ are not the correct bosonic components of the chiral
multiplets of the low-energy $\mc{N}=1$ 4D supergravity. In fact, the
K\"ahler moduli are given by the $T_i=\tau_i + \I\, b_i$ which are related to the $t^i$ as follows:
$\tau_i$ is the Einstein-frame volume (in units of
$l_s$) of the divisor $D_i\in H_4(Y_3,\mbb{Z})$ and reads:
\be
 \tau_i=\frac 12\int_{D_i}\imath^* J\wedge \imath^*J
  = \frac 12\int_{Y_3} \hat{D}_i\wedge J\wedge J
  =\frac{\partial \vo}{\partial t^i}
  =\frac 12\, k_{ijk}\, t^j\, t^k\,, \label{defOfTau}
\ee
where $\imath^* J$ is the pullback of $J$ to $D_i$. Its axionic partner $b_i$ is
the component of the Ramond-Ramond 4-form $C_4$ along this cycle:
$\int_{D_i}\imath^* C_4 = b_i$.

The tree-level K\"ahler potential $K_{\rm tree}=-2\ln \vo$
is given as a function of $T_i$ by solving the
equations (\ref{defOfTau}) for the $t^i$ as functions of the $\tau_i = \frac12(T_i
+\overline{T}_i)$, and substituting the result into
eq.~\pref{IlVolume} to evaluate $\vo$.

\subsection{Diagonalisation of the volume form}
\label{sec:diagonalize}

We shall now look for a change of basis $\{\hat D_i \}_{i=1}^{h^{1,1}}$ to  $\{\hat D'_i \}_{i=1}^{h^{1,1}}$ which
brings the volume, or part of it, in a diagonal form. Parameterising this change of basis
in terms of an ($h^{1,1}\times h^{1,1}$)-matrix $a$ of generic integer coefficients $a_i{}^j$,
the new basis elements can be
expressed as $\hat D'_i= a_i{}^j\,\hat D_j$ with corresponding volumes given by:
\be
\tau'_i=\tfrac 12\int_{Y_3} \hat D'_i\wedge J\wedge
 J=\tfrac 12 \,a_i{}^j\,k_{jpq}\,t^p\, t^q \,.
\ee
We now search for vectors $\vec a_i$ $\in\mbb{Z}^{h^{1,1}}$ such that the divisor volumes $\tau'_i$ become
a complete square in terms of the $t^i$. If there exist any such $\vec a_i$, we call the number of linear independent solutions $n_s\le h^{1,1}$. We take these solutions as the first $n_s$ rows of $a$, while the remaining $\vec a_j$, with $n_s<j\le h^{1,1}(Y_3)$, give the completion of $a$ to a full rank matrix. In the case $n_s=h^{1,1}$ the volume form can be completely diagonalised, resulting in a
`strong' Swiss-cheese Calabi-Yau, while for $n_s< h^{1,1}$ we shall simply speak of `weak' Swiss-cheese manifolds.

In order to find such solutions, we consider the symmetric matrices:
\be
(A_i)_{pq}:= \frac 12 \,a_{i}{}^j\,k_{jpq}\,,
\ee
which admit a similarity transformation to the diagonal matrices $A'_i$:
\be
A'_i=\left(\begin{array}{ccccc}
\lambda_{i1}  & \cdots & \cdots & \cdots &   0\\
\vdots  & \ddots & &  &  \vdots\\
\vdots  &  & \lambda_{in} &  & \vdots\\
\vdots  &  &  & \ddots & \vdots\\
0  & \cdots & \cdots & \cdots & \lambda_{i h^{1,1}}
\end{array}\right)\,.
\ee
The divisor volume $\tau'_i= (A_i)_{pq}\,t^p\, t^q$ becomes a complete square only if the matrix $A_i$ has just one non-zero eigenvalue $\lambda_{in}$ plus ($h^{1,1} -1$) vanishing
eigenvalues $\lambda_{im}$ for all $m\neq n$.

Given that the eigenvalues of $A_i$ are functions, or more precisely sections, of the $h^{1,1}$ integer vectors $\vec a_i$, we have to find a solution to:
\be\label{eq:eigenvalue}
\lambda_m\!\!\left(\vec a_{i}\right)=\lambda_{im}=0\quad\forall\, m\ne n\quad
\text{and}\quad \lambda_n\!\!\left(\vec a_i\right)=\lambda_{in}\ne 0\,.
\ee

\noindent If there exist any solution to~\eqref{eq:eigenvalue}, we can have two situations:
\begin{enumerate}
\item
The inner product between the eigenvector $\vec{v}_{in}$,
corresponding to the only non-vanishing eigenvalue $\lambda_{in}\ne 0$, and $\vec a_i$ does not vanish:
\be
\langle \vec{a}_{i},\vec{v}_{in}\rangle:= a_{i}{}^j\,({\vec{v}_{in}})_j\ne0\,.
\ee
In this case we can expand $\vec a_i$ as a sum over $\vec{v}_{in}$ and
its normal space $\{\vec{v}_{im}\}_{m\ne n}^{h^{1,1}}$ corresponding to the zero eigenvalues:
\be
\vec{a}_{i}=\frac{\langle \vec{a}_{i},\vec{v}_{in}\rangle}
{\langle \vec{v}_{in},\vec{v}_{in}\rangle}\,\vec{v}_{in}
+\sum_{r\ne n}^{h^{1,1}}c_r\,\vec{v}_{ir}\,.
\ee
Therefore we can rewrite the intersection polynomial for every solution $i$ in the following form:
\be
I_3=\lambda_{in}\frac{\langle \vec{a}_i,\,\vec{v}_{in}\rangle^2}
{\langle \vec{v}_{in},\vec{v}_{in}\rangle}
\,D'_{i}\, D'_{i}\, D'_{i}+ k'_{rkl}\,D'_r D'_k D'_l\,,
\ee
with $D'_i= a_{i}{}^j\, D_j$ and
$D'_r= ({\vec{v}_{ir}})^j \, D_j$,
where $j$ runs from 1 to $h^{1,1}$, but $r,\,k$ and $l$ omit $n$.
If $n_s>1$ we see from:
\be
(\vec{v}_{kn})_p\propto (A_k)_{pq}\,a_{i}{}^q
=\tfrac 12\, a_{k}{}^j\,k_{jpq}\,a_{i}{}^q
= (A_i)_{jp}\,a_{k}{}^j\propto (\vec{v}_{in})_p\,,
\ee
that $\vec{a}_i$ has to be normal to all $\vec{v}_{kn}$ with $k\ne i$.
Hence we obtain a basis in which the intersection polynomial further simplifies to:
\be
I_3=\sum_{i=1}^{n_s} \lambda_{i n}\frac{\langle \vec{a}_i,\,\vec{v}_{i n}\rangle^2}
{\langle\vec{v}_{i n} ,\,\vec{v}_{i n}\rangle}
\,D'_i\, D'_i\,D'_i+ \sum_{p,q,r=1}^{h^{1,1}-n_s} k'_{pqr}\,D'_p\, D'_q \,D'_r\,,
\label{eq:dia-form-gen}
\ee
where now each $D'_p$, $p=1,\,\ldots,\,(h^{1,1}-n_s)$, is generated by
a basis element of the space normal to the linear span of the $\vec{v}_{i n}$ with $i=1,\,\ldots,\,n_s$.
We wrote out the sums explicitly to indicate that we sum also over $i$, in contrast to above.

\item
In the case when $\langle \vec{a}_i,\vec{v}_{i n}\rangle=0$, $ \vec{a}_i$ belongs to the normal space to $\vec{v}_{i n}$. Hence the intersection polynomial takes the form:
\be
I_3=\lambda_{in}\langle \vec{v}_{in},\vec{v}_{in}\rangle^2
\,D'_i\, D'_n\, D'_n
+ \sum_{p,q,r\ne l ,n}^{h^{1,1}} k'_{pqr}\,D'_p\,  D'_q\,  D'_r\,,
\label{eq:dia-form-gen-K3}
\ee
where $D'_i= a_i{}^j\, D_j\propto  (\vec{v}_{i l})^j\, D_j$, $D'_n= (\vec{v}_{in})^j\, D_j$, and each $D'_p$, with $n\ne p\ne l$,
is generated by basis elements
of the space normal to the linear span of $\vec{a}_i\propto  \vec{v}_{il}$ and $\vec{v}_{in}$.
For $n_s>1$ we may obtain combinations of \eqref{eq:dia-form-gen} and \eqref{eq:dia-form-gen-K3}.

We notice that if the divisor $D'_i$ is numerically effective then the linearity of the intersection form in $D'_i$, together with
the fact that $D'_i\cdot \,c_2(Y_3) >0$,
implies that the Calabi-Yau is a fibration over $\mbb{P}^1$ with generic fibre given by a K3 surface \cite{Oguiso}.
\end{enumerate}
From this discussion, we conclude that for a divisor $D$ with\footnote{Note that the curve $D\cap D$ may not be effective.}:
\be
\tau_D\propto {\rm Vol}\left(D\cap D\right)^2=\left(\int_{D\cap D} J\right)^2,
\label{eq:diagDiv}
\ee
we can find a basis of divisors such that $I_3=k_D\, D\cdot D\cdot D +...$
where the ellipsis denote intersections independent of $D$.
In this basis the constant of proportionality in (\ref{eq:diagDiv}) becomes:
\be
{\rm Vol}(D\cap D)^2 =2 \, k_D {\rm Vol}(D)\,.
\label{nddP}
\ee
This is the condition that defines what we called a `diagonal' four-cycle.
Moreover, if not obstructed by the K\"ahler cone, it implies the existence of a decoupling limit for $D$ where:
\be
\lim_{J\rightarrow J_0}  {\rm Vol}(D)=0\ne {\rm Vol}(D)\big|_J\simeq \mc{O}(1)\,,
\quad\textmd{and}\quad
\lim_{J\rightarrow J_0} {\rm Vol}(Y_3)\simeq
{\rm Vol}(Y_3)\big|_J\gg\mc{O}(1)\,.
\ee

\section{$D$-terms and shrinking cycles}
\label{sec:d-terms}

In this section we shall study the interplay between a diagonal
structure of the Calabi-Yau volume and the requirement of having
vanishing Fayet-Iliopoulos terms, showing that `diagonal' del Pezzo divisors
are always forced to shrink to zero size.

\subsection{Fluxed $D$-branes and Fayet-Iliopoulos terms}

A very important ingredient of type IIB flux compactifications
is the presence of space-time filling $D7$-branes which can wrap
an internal divisor $D_i$.
Each $D7$-brane comes along with a
$U(1)$ gauge theory that lives on its eight-dimensional world-volume $\mc{W}$.
The brane is embedded in the ten-dimensional space-time manifold
$\mc{X}_{10}=\mathbb{R}^{3,1}\times Y_3$ via the map $\imath:\mc{W}\hookrightarrow
\mc{X}_{10}$.

Furthermore, the $D7$-brane wrapping the divisor $D_i$ can be fluxed,
in the sense that we can turn on a generic internal magnetic flux $F_i$.
The flux can be expanded as \cite{Jockers:2004yj}:
\be
F_i= f_i^k\,\imath^* {D}_k+ f_i^a\,\omega_a,
\ee
with ${D}_k \in H^{1,1}(X)$ and $\omega_a\in {H^{1,1}}^\bot(D_i)$ where $H^{1,1}(D_i)=\imath^* H^{1,1}(X)\oplus {H^{1,1}}^\bot(D_i)$ and $\imath^*$ denotes the pullback operation.
${H^{1,1}}^\bot(D_i)$ is the space of (1,1)-forms which are defined on the
world-volume of the $D7$-brane wrapping $D_i$
such that $\int_{D_i} \imath^* D_j\wedge\omega_a=0$ $\forall \,j$ and $a$.

This gauge flux on $D_i$ induces a $U(1)$-charge $q_{ij}$ for the K\"ahler modulus $T_j$
which takes the form:
\be
q_{ij}= \int_{D_i} \imath^* {D}_j \wedge (F_i-\imath^* B)
=\tilde{f}_i^k \int_{Y_3} {D}_i\wedge{D}_j \wedge {D}_k = \tilde{f}_i^k k_{ijk},
\ee
where we have absorbed the $B$-field coefficients in $\tilde{f}_i^k$.
The K\"ahler modulus $T_{U(1)_i}$ which gets coupled to the $U(1)$ gauge boson living on $D_i$,
is in general a combination of divisors corresponding to the four-cycle Poincar\'e dual
to the two-cycle supporting the magnetic flux.
Due to this coupling, the $U(1)$ gauge boson becomes massive via the St\"{u}ckelberg mechanism
by eating the axion $a_{U(1)_i}=\text{Im}(T_{U(1)_i})$, but $\tau_{U(1)_i}=\text{Re}(T_{U(1)_i})$ remains as a light modulus
in the effective field theory and gives rise
to a moduli-dependent Fayet-Iliopoulos (FI) term $\xi_i$ which looks like \cite{Jockers:2005zy,Haack:2006cy,Dine:1987xk}:
\be
\xi_i=\frac{1}{4\pi\vo}\int_{D_i} \imath^* J\wedge (F_i-\imath^* B)
=\frac{1}{4\pi\vo}\,t^j \tilde{f}_i^k k_{ijk}= \frac{q_{ij}}{4\pi}\frac{t^j}{\vo}
=-\frac{q_{ij}}{4\pi}\frac{\partial K}{\partial \tau_j}.
\label{FI}
\ee
Therefore we end up with a $D$-term scalar potential of the form $V_D=\frac{g_i^2}{2}\,\xi_i^2$
where the gauge coupling constant $g_i$ is given by (considering the dilaton $\phi$ fixed at its VEV: $e^{\phi}=g_s$):
\be
\frac{2\pi}{g_i^2} = \text{Re}(T_i) - \frac{\tilde{f}_i^j q_{ij}}{2 g_s}.
\label{g-2}
\ee
Including also the presence of unnormalised charged matter fields $\varphi_j$ (open string states)
with corresponding $U(1)$ charges given by $c_{ij}$,
the resulting $D$-term potential looks like:
\be
V_D = \frac{g_i^2}{2} \left( \sum_j c_{ij} \varphi_j \frac{\partial K}{\partial \varphi_j} -\xi_i\right)^2
=\frac{2\pi g_s}{2\tau_i g_s- \tilde{f}_i^j q_{ij}} \left( \sum_j c_{ij} \varphi_j \frac{\partial K}{\partial \varphi_j}
+\frac{q_{ij}}{4\pi}\frac{\partial K}{\partial \tau_j}\right)^2.
\label{VD}
\ee

\subsection{Shrinking divisors}

The $D$-term potential is particular importance for the stabilisation of
the four-cycle $\tau_i$ which supports a Standard Model-like
brane construction with chiral matter given that
this modulus cannot be fixed by non-perturbative effects \cite{Blumenhagen:2007sm}
\footnote{This might not be the case for compactifications with $h^{1,1}_-\neq 0$ \cite{Grimm:2011dj}.}.
In fact, the presence of chiral matter with vanishing VEVs generically
kills the instanton contribution to the superpotential.

Other possible contributions to fix $\tau_i$ are string loop
corrections. However, the $D$-terms are the leading effect in a large volume expansion since they scale as $\vo^{-2}$ whereas the string loop
corrections behave as $\vo^{-3}\tau_i^{-1/2}$.

Moreover, the standard LVS potential generated by non-perturbative and $\alpha'$
corrections, behaves like $\vo^{-3}$.
Hence the $D$-term potential has to cancel (at least at leading order up
to $\mc{O}(\vo^{-3})$) in order not to develop a run-away direction for the volume mode.
In the absence of Standard Model singlets which can acquire non-zero VEVs
and for vanishing VEVs of visible sector fields,
the supersymmetric requirement of having $V_D=0$ implies $\xi=0$.
We shall now show that in the case of a `diagonal' del Pezzo
this requirement forces this cycle to shrink to zero size~\cite{Blumenhagen:2008zz}.

First of all we impose the requirement of having no chiral intersections
between the visible $D7$-stack wrapped around $D_i$ and
the ED3 instanton wrapped around a generic rigid divisor $D_{E3}$,
which is needed to generate the non-perturbative effects of the LVS:
\be
\langle \Gamma_i, \Gamma_{E3} \rangle = \int_{Y_3} {D}_i \wedge {D}_{E3} \wedge (F_i-B)
= \tilde{f}^j_{i} k_{ij E3}= q_{i E3}=0,
\ee
where the total world-volume flux on the instanton is zero: $\mc{F}_{E3}= F_{E3}-B=0$
\footnote{We recall that the cycle wrapped by the ED3 is a rigid \textit{non-Spin} four-cycle,
and so we always need to turn on a half integer flux in order to cancel the
Freed-Witten anomalies: $F_{E3}=(1/2){D}_{E3}$. This flux can then be compensated by the $B$-field.}.
This requirement simplifies the expression (\ref{FI}) for the FI term to:
\be
\xi_i= \frac{1}{4\pi\vo}\sum_{j\neq E3} q_{ij}t^j
=-\frac{1}{4\pi}\sum_{j\neq E3}q_{ij}\frac{\partial K}{\partial \tau_j}.
\label{FIsimpl}
\ee
This expression further simplifies if $D_i$ is a `diagonal' del Pezzo since
its defining relation (\ref{nddP}) implies that $k_{ijk}\neq 0$ only if $i=j=k$,
or, in terms of $U(1)$-charges, $q_{ij}\neq 0$ only if $i=j$. Hence (\ref{FIsimpl})
becomes:
\be
\xi_i= \frac{q_{ii}t^i}{4\pi\vo}
= c_i \sqrt{\tau_i},
\label{FInddP}
\ee
where:
\be
c_i = \frac{\tilde{f}_{i}^i}{4\pi\vo} \sqrt{2 k_{iii}}\neq 0.
\ee
Therefore if we do not want to lead the system to any decompactification limit corresponding to $\vo \to \infty$,
$\xi_i =0$ implies $\tau'_i\to 0$. This result is in agreement with previous explicit Calabi-Yau
examples. In fact, the authors of \cite{Blumenhagen:2008zz} presented an example
with a diagonalisable volume which is suitable to realise the LVS but
where all the rigid divisors are dP$_8$ four-cycles which
are forced to shrink to a point by the $D$-terms. Similar situations
where found in \cite{Collinucci:2008sq} where the authors
presented several Calabi-Yau examples with a `strong' Swiss cheese structure. However, only some of the diagonal divisor were del Pezzo and could be contracted without interfering with the volume of the other divisors.
All cases were promising to realise the LVS but there were always one or more four-cycles
shrinking to zero size due to the supersymmetric condition $D=0$.

The shrinking of a cycle is generically considered as a problem because
it is hard to have control over $\alpha'$ and
quantum corrections at the singular regime,
while the effective field theory in the geometric regime is much more under control.
However if one does not care that much about this control issue,
then the shrinking of a cycle can even be welcome in order to reproduce the nice
phenomenological features of models at the quiver locus \cite{Conlon:2008wa}.
Some of these constructions with fractional branes at singularities allow
the sequestering of the visible sector with a hierarchy between $M_{soft}$ and $m_{3/2}$,
$M_{soft}\sim m_{3/2}^2/M_P$, which might lead naturally to both GUT-theories
and TeV-scale SUSY \cite{Blumenhagen:2009gk}.

Therefore we showed how a `diagonal' del Pezzo divisor is the best candidate
either to support the non-perturbative effects of the LVS (in this case no FI-term
gets generated) or to lead the system to the quiver locus.

\subsection{$D$-terms and geometric regime}

We have argued that every time the branes involved in the
GUT construction wrap divisors appearing in the volume in a diagonal manner, then
the vanishing FI term condition, always forces some cycles to
shrink to zero size.

Let us now try to discuss possible ways to avoid this shrinking limit so
that the visible sector can be built via magnetised branes wrapping cycles in the geometric regime:
\begin{enumerate}
\item Consider brane constructions with Standard Model singlets (like a right handed sneutrino)
which can get a non-vanishing VEV cancelling the FI term:
$\langle|\varphi|\rangle\sim\sqrt{\xi}$.
In this way the FI term is not forced to zero anymore and the
cycle supporting the visible sector could then be fixed using
string loop corrections to the K\"ahler potential \cite{Cicoli:2008va}.

\item
Wrap around the shrinking cycle the instanton giving rise to the
top Yukawa coupling \cite{Blumenhagen:2008zz,Cvetic:2010rq}\nocite{Cvetic:2010ky}. More precisely, an $O(1)$ instanton wrapping a
rigid cycle invariant under
the orientifold action and with the following chiral intersections with the GUT stack $D7_a$
and the intersecting stack $D7_b$:
\be
I_{a,inst}=1,\qquad I_{b,inst}=-1,
\ee
generates perturbatively forbidden top-Yukawa couplings:
\be
Y_{\alpha\beta}\textbf{10}^{\alpha}\textbf{10}^{\beta}\textbf{5}_{H}\,
e^{-\tau_{inst}},
\ee
which are exponentially suppressed, and so very
far from their $\mc{O}(1)$ phenomenological values. This is
one of the main reasons for considering F-theory instead of type
IIB GUT constructions. However if the instanton cycle shrinks to zero size,
then the top-Yukawa coupling becomes phenomenologically viable:
$e^{-\tau_{inst}}\to 1$ for $\tau_{inst}\to 0$.

\item Consider Calabi-Yau three-folds without a diagonal structure for the cycles
supporting the branes involved in the GUT construction, but still
with a rigid divisor which is orthogonal to all these four-cycles and
appears in the volume in a diagonal way. This further cycle is
needed to get the LVS.

The authors of \cite{Blumenhagen:2008zz} presented a case where all rigid divisors are dP$_9$,
hence they are not contractible to a point \footnote{
They also proposed to use models with intersecting dP$_8$
that do not allow for a diagonal structure of the volume.}. The volume does not have a diagonal
structure, and so there is no shrinking cycle due to $D$-terms.
However the LVS could not be realised due to the absence of a `diagonal' del Pezzo.

\item Place the GUT theory on the
K3 fibre which does not appear in the volume
in a diagonal manner, and so the $D$-terms might not force any cycle to contract to a point.
However the K3 fibre deformations should be lifted
in order not to get unwanted matter in the adjoint
representation. Moreover, contrary to previous studies \cite{Cicoli:2008va},
the K3 fibre should be stabilised at a relatively small value in order not to get
an exponentially small value for the gauge coupling of the visible sector.
\end{enumerate}

\section{The search for suitable K3 fibrations}
\label{sec:k3search}

In this section we describe how to search for Calabi-Yau hypersurfaces that admit a K3 fibration
and a `weak' Swiss-cheese type volume structure. As we have already pointed out,
this kind of manifolds are needed to realise particular versions of the LVS which
are very interesting for their applications to cosmology \cite{Cicoli:2008gp,Burgess:2010bz,Cicoli:2010ha,Anguelova:2009ht}
and particle phenomenology \cite{Cicoli:2011yh,Cicoli:2011yy}.

The first input parameter for our search is the number of K\"ahler moduli $h^{1,1}$ of the Calabi-Yau $Y_3$.
Moduli stabilisation requires already the presence of a third modulus, a `diagonal' del Pezzo,
in addition to the two moduli controlling the volume of the base and the K3-fibre.
We focus however on $h^{1,1}=4$, since, we are interested in finding
a second rigid divisor for a chiral $D$-brane configuration.

Furthermore, we want to use toric techniques to control all the K\"ahler moduli
(see~\cite{Fulton:1993,Oda:1987} for a mathematical description of the notation
and the necessary tools, or~\cite{Skarke:1998yk,Bouchard:2007ik,Kreuzer:2006ax,Hori:2003ic,Braun:2010ff}
for more physical introductions to this topic).
Therefore, we take into account only models that have all
their Picard generators induced from the toric ambient four-fold $X_4$. In the heterotic literature,
these manifolds are called favourable~\cite{Anderson:2008uw}.

With these input data, we perform our search not starting
from the list of K3-fibred toric Calabi-Yau hypersurfaces~\cite{Avram:1996pj},
but, for convenience and statistical reasons, we prefer to
scan the list of four-dimensional reflexive lattice polytopes~\cite{Kreuzer:2000xy,cydata}
finding 1185 polytopes with $h^{1,1}=4\le\textmd{Pic}_{X_4}$ \footnote{$\textmd{Pic}_{X_4}\ge 4$ is only
a necessary condition for inducing four primitive divisor classes
on the three-fold from the toric ambient space. The reason is that a toric divisor,
which corresponds to a real interior point of a maximal dimensional facet of the polytope,
would not intersect the Calabi-Yau hypersurface.}.
Out of these Calabi-Yau hypersurfaces
we pick those that exhibit a reflexive section of co-dimension one. Since,
these are the models that, in the case of a compatible triangulation, admit a K3 fibration structure.
In order to check for a reflexive section, we use the Package to Analyse Lattice Polytopes PALP~\cite{Kreuzer:2002uu}
which can scan for such a section up to co-dimension three.
Imposing this selection criterion we are left with only 650 polytopes.

After having a list of relevant K3-fibred Calabi-Yau three-folds,
we seek for those that have a volume form of `weak' Swiss-cheese type.
To be more specific, we demand the presence of a rigid divisor $D$
which is a `diagonal' del Pezzo~\eqref{eq:diagDiv} given that this is the particular
kind of four-cycle that can support the non-perturbative effects needed to realise the
LVS ~\cite{Balasubramanian:2005zx,Cicoli:2008va}. In order to check this property,
we have to construct the toric varieties corresponding to the lattice polytopes
and calculate the properties of the toric divisors intersecting the Calabi-Yau hypersurface.

The toric ambient space can be obtained by triangulating the polytope.
Hence, in the case when the triangulation of a polytope is not unique,
we find more than just one toric variety for the polytope. This may not necessarily result in different Calabi-Yau three-folds, for simplicity however, we will treat all these hypersurfaces as different varieties.
We work out the triangulations using an extended version of PALP~\cite{Braun:2011ik}
which calculates also the Mori cone of the toric variety and, with the help of SINGULAR \cite{SINGULAR},
the intersection ring of the toric divisors on the Calabi-Yau. With the intersection numbers at hands,
we then obtain the topological numbers of the toric divisors on the hypersurface.

Imposing one by one the above constraints, we find 540 lattice polytopes
with at least one del Pezzo surface and 228 with at least two.
As we have already stressed above, a second toric del Pezzo divisor
might be needed in order to realise chiral $D$-brane constructions
within a LVS with stabilised closed string moduli.
Finally, out of the 540 polytopes, 158 have at least one triangulation
that admits a K3 fibration with a `diagonal' del Pezzo.
The list of these polytopes is given in appendix~\ref{sec:the-list}.

In the next two sections we shall work out in detail
some illustrative examples chosen from these 158
polytopes. We shall start from the case of simplicial ones
where all the del Pezzo divisors are `diagonal', and then
move to discuss the case of non-simplicial polytopes
with the interesting emergence of `non-diagonal' del Pezzo four-cycles.

\section{K3 fibrations from simplicial polytopes}
\label{sec:simp-polytopes}

The first class of examples that we consider, are Calabi-Yau hypersurfaces
embedded in toric ambient spaces which are given by simplicial lattice polytopes.
Here, simplicial means that the number of vertices of the polytope equals the dimension plus one.
For the corresponding toric manifolds all but one divisor classes
come from resolutions of singularities of the toric variety.

Out of the eleven four-dimensional, simplicial, reflexive lattice polytopes with $h^{1,1}(Y_3)=4$,
ten have more than eight points, including the origin, such that $\textmd{Pic}_{X_4}\ge4$.
However, only eight of them induce four primitive divisor classes from the toric four-fold on the
Calabi-Yau hypersurface. Moreover, only six of them have a reflexive section of co-dimension one and five a triangulation
which respects the fibration structure.
The corresponding spaces have all the same structure.

The Calabi-Yau $\mbb{P}^4_{[1,1,2,2,2]}(8)/\mbb{Z}_2\!\!:\!0\,1\,1\,0\,0$ T1,
where T1 stands for the `first' triangulation,
will be the archetype example that we shall analyse in detail below. In this case
we shall be able to find a basis in which the intersection polynomial completely diagonalises.
The other Calabi-Yau manifolds of this type are the three further triangulations of
$\mbb P_{[1,1,2,2,2]}(8)/\mbb Z_2\!\!:\!0\,1\,1\,0\,0$
with a K3 fibration structure, one triangulation of
$\mbb P_{[1,1,2,2,2]}(8)/\mbb Z_2\!\!:\!0\,1\,1\,1\,1$,
four triangulations of $\mbb P_{[1,1,2,2,6]}(12)/\mbb Z_2\!\!:\!0\,1\,1\,0\,0$,
one triangulation of $\mbb P_{[1,1,2,2,6]}(12)/\mbb Z_2\!\!:\!0\,1\,1\,1\,1$,
and one, out of the four, triangulation of $\mbb P_{[1,1,2,2,6]}(12)/\mbb Z_2\!\!:\!1\,0\,0\,0\,1$
that admits a K3 fibration.
They all share very similar intersection structures, and so, like in the illustrative
$\mbb{P}^4_{[1,1,2,2,2]}(8)/\mbb Z_2\!\!:\!0\,1\,1\,0\,0$ T1 case,
their volume can be completely diagonalised.

\subsection{The Exemplar: $\mbb P^4_{[1,1,2,2,2]}(8)/\mbb Z_2\!\!:\!0\,1\,1\,0\,0$ T1}
\label{sec:theexemplar}

The toric ambient space $\mbb{P}^4_{[1,1,2,2,2]}(8)/\mbb Z_2\!\!:\!0\,1\,1\,0\,0$
can be defined by homogeneous coordinates and their equivalence relations.
The weight matrix for the resolved variety reads:
\be
\begin{array}{|c|c|c|c|c|c|c|c||c|}
\hline z_1 & z_2 & z_3 & z_4 & z_5 & z_6 & z_7 & z_8 & D_\textmd{H} \tabularnewline \hline \hline
2 & 1 & 0 & 0 & 2 & 2 & 0 & 1 & 8\tabularnewline\hline
1 & 0 & 0 & 0 & 2 & 2 & 2 & 1 & 8\tabularnewline\hline
1 & 1 & 0 & 2 & 2 & 2 & 0 & 0 & 8\tabularnewline\hline
1 & 0 & 1 & 0 & 1 & 1 & 0 & 0 & 4\tabularnewline\hline
\end{array}\label{eq:wm-11222}\,.
\ee
Here, the $z_i$ are the homogeneous coordinates and $D_\textmd{H}$ stands for the multi-degree of the hypersurface (divisor) which, in the Calabi-Yau case, is just the sum of the weights in a row.
The surface of the reflexive lattice polytope encoding this weights exhibits several maximal triangulations.
Given that we may construct a fan from each triangulation, we obtain more than just one toric variety for this weight matrix.
In the following, we stick to the `first' triangulation, T1, out of the four for this polytope.
The triangulation and the fan, respectively, can be encoded in the Stanley-Reisner (SR) ideal, for which we obtain:
\be
 {\rm SR}=\{z_1 z_8 ,\, z_2 z_8,\,z_1 z_2,\, z_1 z_3,\, z_7 z_8,\, z_3 z_7,\, z_2 z_6
z_4 z_5,\, z_3 z_4 z_5 z_6,\, z_4 z_5 z_6 z_7\}\,.
\ee
The SR-ideal is the set of homogeneous coordinates that must not vanish simultaneously.
From this and the linear equivalence relations given by the weight matrix we calculate
the triple intersection numbers for a basis of divisor classes of the toric variety on the Calabi-Yau.
Choosing the divisors $D_i=\{z_i=0\}$, $i=1,...,4$ as a basis\footnote{For ease of read, throughout all the paper we arranged the homogeneous coordinates in such a way that the first four are the (relevant) basis even if in general this is not the case.},
their triple intersections can be written in terms of the following polynomial:
\bea
I_3&=&-24 D_4^3 - 8 D_3^2 D_4 - 4 D_2 D_4^2 +4 D_2 D_3 D_4
-2 D_2 D_3^2 -4 D_2^2 D_4 \nonumber \\
&& +2 D_2^2 D_3 -2 D_2^3 + 4 D_1 D_4^2\,.
\eea
The prefactors denote the intersection numbers of the respective divisors.
Moreover, the generators $\mc{C}_i$ of the Mori cone,
i.e.~the cone of effective curves, of the toric variety are:
\be
 \int_{\mc{C}_i} D_j=\left(\begin{array}{cccc}1 & 1 & 0 & 0
 \\ 0 & -1 & 1 & -1 \\ 0 & 1 & -2 & 0 \\ 0 & 0 & 1 & 2\end{array}\right)\,.
\ee
The K\"ahler form can be expanded in the previous basis as:
\be
J=\sum_{i=1}^4 t_i\,D_i\,,
\ee
where we wrote down the sum explicitly and
lowered the index of the $t$'s for ease of read.
It should however be understood that the $t$'s and the $\tau$'s are dual variables.
We shall however perform a change of basis expanding $J$ in terms of
the generators $\Gamma$'s of the K\"ahler cone, i.e.~the dual to the Mori cone,
which in this basis becomes very simple:
\be\label{eq:KaehlerformGenerator}
J=\sum_{i=1}^4 r_i\,\Gamma_i \quad\textmd{with}\quad r_i>0.
\ee
The generators of the K\"ahler cone look like:
\bea
&&\Gamma_1= D_1\,,\quad \Gamma_2=4\,D_1 -4\,D_2 -2\,D_3 +D_4\,,\\\nonumber
&&\Gamma_3=3\,D_1 -3\,D_2 -2\,D_3 +D_4\,,\quad
\,\Gamma_4=2\,D_1 -2\,D_2 -1\,D_3+D_4\,.
\eea
With the K\"ahler form we can calculate the volumes of the divisors
$D_i$, $i=1,...,4$ which turn out to be:
\bea
\tau_1 &=& \tfrac{1}{2}\int_{D_1}J\wedge J=2\,(r_2 + r_3 + r_4)^2\,,\quad
\tau_2 = \tau_1 - (r_3 + r_4)^2\,,\nonumber\\
\tau_3 &=& \tau_1 - \tau_2 - r_4^2\,,\qquad
\tau_4 = 2\left(r_1\sqrt{2\tau_1} +2\tau_2+ \tau_3-\tau_1\right)\nonumber\,.
\eea
The volume of the Calabi-Yau manifold is given by:
\bea
\vo & = &\frac 16 \int_{Y_3}J\wedge J\wedge J \nonumber \\
&=& \left[r_1
 +\frac 23\left(
r_2 + r_3 + r_4\right)\right] 2(r_2 + r_3 + r_4)^2 - \frac 13\, (r_3 + r_4)^3 -\frac 13\, r_4^3 \\
&=&  \frac{\sqrt{\tau_1}}{6\sqrt{2}}\left(10 \tau_1-12\tau_2-6\tau_3+3\tau_4 \right)
-\frac 13\, (\tau_1 - \tau_2 -\tau_3)^{3/2} - \frac 13\, (\tau_1 - \tau_2)^{3/2}. \nonumber
\label{Ilvolume}
\eea
The form of the volume (\ref{Ilvolume}) suggests that we can find a `diagonal' basis for the divisor classes.
Following the algorithm presented in section~\ref{sec:diagonalize}, we define:\footnote{The definition
of $D_{K3}$ will become clearer later on.}
\bea
\label{eq:Exemplardiabasis}
D_a & := &  D_1 =:D_{K3}\,, \quad
D_b  :=   10 D_1-12 D_2-6 D_3 + 3 D_4\,,\nonumber \\
D_c & := & D_1 - D_2 - D_3=D_8 \,,\quad
D_d  :=   D_1 - D_2\,.
\eea
In this basis, the intersection polynomial takes the form:
\be
 I_3 = 36\,D_a\,D_b^2+2\,D_c^3+2\,D_d^3,
\ee
showing that \eqref{eq:Exemplardiabasis} is indeed a diagonal basis.
The volumes of the diagonal divisors are:
\bea
\tau_a &=& \tau_1 = 2\,(r_2 + r_3 + r_4)^2\,,\quad
\tau_b = 10 \tau_1-12\tau_2-6\tau_3+3\tau_4
= \left[r_1
 +\frac 23\left(
r_2 + r_3 + r_4\right)\right] 6\sqrt{2\tau_1} \nonumber \\
\tau_c &=& \tau_1 - \tau_2 -\tau_3=r_4^2\,,\qquad
\tau_d = \tau_1 - \tau_2=(r_3 + r_4)^2\,.
\eea
Finally, we can rewrite the Calabi-Yau volume in terms of the above expressions:
\be
\vo  = \alpha\left(\tau_b\sqrt{\tau_a}
-\beta\, \tau_c^{3/2} - \beta\, \tau_d^{3/2}\right),
\ee
where $\alpha= 1/\left(6\sqrt{2}\right)$ and $\beta = 2\sqrt{2}/3$.
We point out that we managed to obtain a K3-fibred Calabi-Yau three-fold
of `strong' Swiss-cheese type that is of the same form of the
compactification manifolds used in \cite{Cicoli:2008gp,Burgess:2010bz,Cicoli:2009zh}
for very promising cosmological applications.
Moreover, in the limit of small $\tau_c$ and $\tau_d$ (for small $r_3$ and $r_4$),
the overall volume is controlled by the two large cycles $\tau_a$ and $\tau_b$:
\be
\vo  \simeq \alpha\,\tau_b\sqrt{\tau_a} \simeq r_1 \tau_1 + \frac{\sqrt{2}}{3} \,\tau_1^{3/2},
\ee
where, as we shall see in the next section,
$\tau_a=\tau_1$ gives the volume of the K3 fibre while $r_1$
is the volume of the $\mbb{P}^1$ base of the fibration which
is embedded into the Calabi-Yau by one of the four $\mbb P^1$'s of $D_4\cap D_7$.

It is interesting to notice that in the large volume limit $r_1\gg r_2$ $\Leftrightarrow$
$r_1\gg \sqrt{\tau_1}$, the characteristic size of the two-dimensional base is
much bigger than the size of the four dimensions of the fibre. This anisotropic shape
was used in \cite{Cicoli:2011yh,Cicoli:2011yy} for very interesting phenomenological
applications.

\subsubsection{Divisor analysis}

From the weight matrix (\ref{eq:wm-11222}), we observe that there are five rigid divisors
$D_2$, $D_3$, $D_4$, $D_7$ and $D_8$, i.e.~the respective normal bundles do not have any non-vanishing global section on these divisors.
A surface that is shrinkable or, in other words, that admits a blow-down map,
should be an exceptional divisor which is always rigid.
Therefore from now on we shall focus on these five rigid divisors
and determine their geometry, trying to understand which of them is a del Pezzo surface.
Notice that none of these divisors can be the K3 fibre, since the fibre must have a trivial normal bundle.

Using the adjunction formula in the form of \cite{Denef:2008wq}:
\be
 c(S)=\frac{\prod_i^8(1+D_i)}{(1+D_\textmd{CY})(1+S)}=1+ c_1(S)+c_2(S)+\ldots \,,
\ee
we can calculate the first and second Chern classes, $c_1$ and $c_2$,
of the tangent bundles of a generic divisor $S$.
From $c(S)$ we obtain the Todd classes, Td$=1+\frac 12 c_1+\frac{1}{12}(c_1^2+c_2)+\ldots$,
that allow us to compute the Euler and holomorphic Euler characteristic of $S$:
\be
\chi(S)=\sum_{i=0}^{2\textmd{dim} (S)}(-1)^i\, b^i=\int_S c_{\textmd{dim} (S)}(S)
\quad\textmd{and}\quad\chi_h(S)=\sum_{i=0}^{\textmd{dim} (S)}(-1)^i\,h^{i,0}=\int_S \textmd{Td}(S)\,,
\ee
with $\textmd{dim} (S)$ denoting the complex dimension of $S$.
The considered divisors have the following intersection structures with the other divisors:
\be
\label{tab:intersectiontable}
\begin{array}{|c|c|c|c|c|c|c|c|c|}
\hline & D_1 & D_2 & D_3 & D_4 & D_5 & D_6 & D_7 & D_8 \\ \hline \hline
D_2^2 & 0 & -2 & 2 & -4 & -2 & -2 & 0 & 0 \\\hline
D_3^2 & 0 & -2 & 0 & -8 & -4 & -4 & 0 & 2 \\\hline
D_4^2 & 4 & -4 & 0 & -24 & -8 & -8 & 0 & 8 \\\hline
D_7^2 & 4 & 4 & 0 & -8 & -8 & -8 & -16 & 0 \\\hline
D_8^2 & 0 & 0 & -2 & -4 & -2 & -2 & 0 & 2 \\\hline
\end{array}\,.
\ee
The divisor $D_8$ has triple self-intersection $D_8^3=2$
and otherwise only vanishing or negative entries in table (\ref{tab:intersectiontable}).
This translates into:
\be
\label{poscurve}
\int_{\mc{C}=D_i\cap S}c_1(S)=\int_{Y_3} D_i\wedge S\wedge(- c_1(\mathcal N_{S|Y_3}))=-S^2\cdot D_i > 0
\qquad \forall \,\mc{C}:\, D_i\neq S \,\wedge\,D_i\cap S\neq\emptyset\,,
\ee
which is a necessary condition for $S$ to be a del Pezzo divisor.
The Euler characteristic and holomorphic Euler number of this divisor are ten and one, respectively.
These are the topological data of a dP$_7$.

We can show algebraically that $D_8$ is a del Pezzo four-cycle by
starting to analyse the toric divisor $D_8$, $\{z_8=0\}$, without the hypersurface equation.
In order to obtain its defining (three-dimensional) fan,
we take the part of the (four-dimensional) fan that is made up by the maximal dimensional cones of the fan
that contain the ray corresponding to $z_8$, and project it along this ray.
The resulting space is $\mbb{P}_{[1,2,1,1]}$ with
$(z_3,z_4,z_5,z_6)\sim (\lambda\,z_3,\lambda^2\,z_4,\lambda\,z_5,\lambda\,z_6)$
if we keep the numbering as it was before the projection.
Setting $z_8=0$ in the hypersurface and restricting to $D_8$
is equivalent to dropping all the monomials with $z_8$ and
setting to one all the homogeneous coordinates $z_j$ that lie with $z_8$ in the SR-ideal, i.e.~$z_j\,z_8\subset \textmd{SR}$.
In toric language this corresponds to taking a section through the dual M-lattice polytope.
Therefore we obtain an equation of degree four in the homogeneous coordinates
$z_3$, $z_4$, $z_5$, and $z_6$.
Given that $\mbb{P}_{[1,2,1,1]}(4)$ is a del Pezzo seven surface,
we do not only know the topological numbers of the intersection of $D_8$
with the Calabi-Yau hypersurface but also its geometry.
For the other rigid surfaces the intersection numbers \eqref{tab:intersectiontable}
do not fulfill \eqref{poscurve}, implying that these divisors cannot be del Pezzo.
Therefore we shall not further analyse them.

As a next step, we get to the K3 divisor.
By computing tables like the one above for the rest of the divisors,
one can easily find that $\int_{D_1} c_1(D_1) \wedge \imath^* D_i=-D_1\cdot D_1 \cdot D_i=0$.
Since we have also that $\int_{D_1}\imath^*c_2(Y_3)>0$, the theorem of~\cite{Oguiso} implies
that this Calabi-Yau three-fold is a K3 fibration over $\mathbb P^1$ with typical fibre $D_1$.
This can be checked by a calculation of the Euler and holomorphic Euler characteristics of $D_1$ which are $24$ and two, as it must be for a K3 surface. Repeating the same analysis as in the case of $D_8$,
we obtain $\mbb{P}^3(4)$ for the intersection of $D_1$ with the hypersurface.

Furthermore, we can also make the projection $\pi$ of the fibration explicit. We will explain the details of the derivation of this map for the next example and give here just the result:
\be
 \pi:\quad [z_1:\ldots:z_8]\mapsto [z_1:z_2\,z_3\,z_8]\in\mathbb P^1\,,
\ee
where $\mathbb P^1$ is the base of the fibration. This fibration has also a section since the hypersurface equation becomes independent of $z_1$, $z_2$, $z_3$ and $z_8$ at the locus $D_4\cap D_7$. One can choose one of the four solutions of $z_5^4=\alpha\, z_6^4$, with some $\alpha\in \mathbb C$, as the embedding of the $\mbb{P}^1$ base of the fibration.

Finally, we look at the shrinkability of these cycles recalling that,
according to Grauter's criterion~\cite{Grauter:1962,Cordova:2009fg},
on a Calabi-Yau only del Pezzo four-cycles can be contracted to a point.
Indeed, taking the limit $r_4\rightarrow 0$ $\Leftrightarrow$ $D_8\to 0$
does not force any of the other cycles to shrink to zero size.
The other rigid divisors are at most contractible to a curve, e.g.~$D_3$ and $D_7$,
without severely influencing the other cycles.

The other triangulations of this polytope which admit a K3 fibration, T2 and T4,
can be related to T1 via flop transitions.
For T2 the properties of $D_7$ and $D_8$ get exchanged with those of $D_2$ and $D_4$.
On the other hand, the last triangulation T4 respects the symmetry of the polytope
and treats $D_2$, $D_8$ and $D_4$, $D_7$ equivalently.
In this case both $D_2$ and $D_8$ become a del Pezzo four-cycle.

\subsubsection{Orientifolding}

Due to the fact that type IIB compactifications give rise to $\mc{N}=1$
supersymmetry in four dimensions only if the internal manifold is
a Calabi-Yau orientifold, we study
what happens to the geometry in the orientifold case.

Depending on the choice of the involution, we have to fix several complex structure moduli
so that the Calabi-Yau hypersurface becomes invariant under the involution.
For an involution $z_i\leftrightarrow -z_i$, the transformed coordinate
must have even powers in the hypersurface equation, and so
all the monomials where the coordinate shows up with odd powers are set to zero.
Therefore we do not consider generic Calabi-Yau hypersurfaces,
but only those that lie on a particular subspace of the complex structure moduli space.

For the Calabi-Yau three-fold under consideration, we shall consider the $\mbb Z_2$-involution $z_5\leftrightarrow -z_5$. Following the algorithm presented in~\cite{diss}, we obtain the following fixed point set of the toric four-fold:
\be
\{\textmd{Fixed}\}\big|_{z_5\leftrightarrow -z_5}=\{D_5\}\,.
\ee
Therefore we end up with one $O7$-plane at the intersection of the Calabi-Yau hypersurface with the toric divisor $D_5$.

\section{K3 fibrations from non-simplicial polytopes}
\label{sec:non-simp-polytopes}

In this section we present two illustrative examples from non-simplicial polytopes.
The first case admits two del Pezzo divisors: one `diagonal' to support the non-perturbative
effects that lead to moduli stabilisation, and one `non-diagonal' for chiral $D$-brane
constructions in the geometric regime.
On the other hand, the second case admits just one `diagonal' del Pezzo four-cycle
but it features an interesting F-theory uplift
in terms of an elliptically fibred Calabi-Yau four-fold.
Another reason why we chose these models is because, besides featuring
at least one `diagonal' del Pezzo divisor,
they also have one homogeneous coordinate with a high multi-degree.
This is an advantage in the case of type IIB model building on Calabi-Yau orientifolds
where there is a generic need of $O$-planes with large divisor classes.

\subsection{A K3 fibration with two del Pezzo divisors}
\label{sec:6.1}

Like in section \ref{sec:theexemplar}, we give the toric ambient variety by its weight matrix:
\be
\begin{array}{|c|c|c|c|c|c|c|c||c|}
\hline z_1 & z_2 & z_3 & z_4 & z_5 & z_6 & z_7 & z_8 & D_\textmd{H} \tabularnewline \hline \hline
1 & 1 & 0 & 0 & 0 & 4 & 1 & 1 & 8\tabularnewline\hline
0 & 1 & 0 & 0 & 1 & 6 & 2 & 2 & 12\tabularnewline\hline
0 & 0 & 1 & 0 & 0 & 2 & 1 & 1 & 5\tabularnewline\hline
0 & 0 & 0 & 1 & 0 & 3 & 1 & 1 & 6\tabularnewline\hline
\end{array}\label{eq:model2dP:weightm}\,,
\ee
and its Stanley-Reisner ideal:
\be
\label{eq:model2dP:sr-ideal}
{\rm SR}=\{z_1 z_2,\,z_2 z_5, \, z_1 z_3,\, z_1 z_4,\, z_4 z_6,\, z_3 z_7 z_8,\, z_5 z_6 z_7 z_8\}\,.
\ee
Choosing the divisors $D_i=\{z_i=0\}$, $i=1,...,4$ as a basis,
we obtain the following intersection structure on the Calabi-Yau hypersurface:
\be
 I_3= 2\,D_1^3 -2\,D_2\,D_3^2 +2\,D_3^3 +3\,D_2\,D_3\,D_4
-2\,D_2\,D_4^2-6\,D_3\,D_4^2 +8\,D_4^3\,.
\ee
Note that in this basis the divisor $D_1$ has a manifest diagonal structure.
The generators $\mc C_j$ of the Mori cone of the four-dimensional toric variety defined via \eqref{eq:model2dP:weightm}
and \eqref{eq:model2dP:sr-ideal}, are:
\be
\label{eq:model2dP:moricone}
\int_{\mc C_j}D_i=\left(\begin{array}{cccc}
-1 & 0 & 0 & 0\\
 3 & 0 & 0 & 2 \\
 0 & 1 & 0 &-2\\
 0 & 0 &-1 & 1\\
 0 & 0 & 3 &-2\end{array}\right)=:\mc{M}_{ji}\,,
\ee
showing explicitly that the polytope is non-simplicial since $j$ runs
from one to five instead of four.
We proceed as in section \ref{sec:theexemplar}
and expand the K\"ahler form in terms of the generators $\Gamma$'s of the K\"ahler cone
which are dual to~\eqref{eq:model2dP:moricone}:
\be
 J=\sum_{j=1}^5 r_j \,\Gamma_j\quad\textmd{with}\quad r_j>0\,,
\ee
and:
\bea
&&
\Gamma_1=D_2\,,\quad
\Gamma_2=2\,D_2 + D_3 + D_4\,,\quad
\Gamma_3=-2\,D_1 +6\, D_2 +2\,D_3 +3\,D_4\\\nonumber
&&
\Gamma_4=-2\,D_1 +6\,D_2 +3\,D_3 +3\,D_4\,,\quad\,
\Gamma_5=6\,D_2 +2\,D_3 +3\,D_4\,.
\eea
The volumes of the divisors $D_i$, $i=1,...,4$, expressed in terms of the K\"ahler parameters $r_j$,
 turn out to be:
\bea
\tau_1 &=& 4\,(r_3 + r_4)^2\,,\qquad \tau_4 = r_1 (r_2 + 3 r_4)\,, \nonumber \\
\tau_2 &=& (r_2+r_3+3\,r_4+r_5)(r_2+5\,r_3+3\,r_4+5\,r_5)-(r_2+3\,r_4)(r_3+r_5)\,, \\
\tau_3 &=& r_1 \left(r_2+5 r_3+3 r_4+5 r_5\right)+\left(r_3+r_5\right) \left(2 r_2+7 r_3+6 r_4+7 r_5\right)\,,\nonumber
\label{taus}
\eea
while the Calabi-Yau volume reads:
\bea
 \vo&=&r_1\tau_2+\tfrac 23 \,r_2^3-\tfrac 13\, \tau_1^{3/2} \\
 &+&\tfrac 13 r_3 [r_2(18\,r_2 + 51\,r_3+108\,r_4 +102\,r_5)
 + r_3(28\,r_3+105 \,r_4+132\,r_5)
 + r_4 (114\,r_4+306\,r_5)] \nonumber \\
 &+&\tfrac 13  r_4[r_2(18\,r_2+54\,r_4+108\,r_5)+r_4(38\,r_4 +162\,r_5)+1710 \,r_5^2] \nonumber\\
 &+& 2\,r_5[r_2(3\,r_2 + 51\,r_5) + 44\,r_5^2]\,.\nonumber
\eea
Given that we are dealing with a non-simplicial K\"ahler cone, the $\Gamma$'s are linearly dependent
even if we have only four K\"ahler moduli. However, in order to be able to write down
the effective field theory, we need to know the independent degrees of freedom which become manifest
if we expand the K\"ahler form in terms of the initial basis $D_i$, $i=1,...,4$:
\be
J=\sum_{i=1}^4 t_i\,D_i\,,\quad\textmd{with}\quad \sum_{i=1}^4\mc{M}_{ji} t_i>0\,.
\ee
The volume of the basis divisor and the Calabi-Yau three-fold as a function of the four linear independent $t$'s looks like:
\bea
\tau_1 &=& t_1^2\,,\qquad \tau_2 =5 t_3 t_4 - (t_3 + t_4)^2\,, \\
\tau_3 &=& t_3^2-3 t_4^2+t_2 \left(3 t_4-2 t_3\right)\,,\quad \tau_4 = (t_2 - 2 t_4) (3 t_3 - 2 t_4)\,, \nonumber
\eea
and:
\bea
\vo & = &t_2\, [5 t_3 t_4 - (t_3 + t_4)^2]\,+\frac 13 \left[t_3^3 - 9\, t_3 \,t_4^2 + 4 \,t_4^3 \right]+ \frac{t_1^3}{3}  \nonumber \\
&=& (t_2 - 2 \,t_4) \tau_2 -\frac 13 (t_3-2 t_4) (\tau_2 + t_3 t_4) - \frac 13\, \tau_1^{3/2}.
\label{Ilvolumee}
\eea
Contrary to the previous simplicial example, we have not been able
to write the volume just in terms of K\"ahler moduli $\tau_i$. Thus
we did not manage to diagonalise the volume form completely, even if the modulus $\tau_1$ is
manifestly diagonal.

As we shall explain in the next section, this manifolds admits two del Pezzo divisors:
$D_1$ which is a dP$_7$ and $D_4$ which is a dP$_1$.
However only $D_1$ turns out to be a `diagonal' del Pezzo, while $D_4$ is a
`non-diagonal' del Pezzo four-cycle which could be very useful for chiral $D$-brane model building.

Let us now show that we can indeed write down the volume form just in terms of the $\tau$'s
in the anisotropic large volume limit used in \cite{Cicoli:2011yh,Cicoli:2011yy}.
We start by taking the limit of small $\tau_1$. As can be seen from (\ref{taus}),
this is equivalent to sending $r_3\to 0$ and $r_4\to 0$. In this limit, the
divisor volumes become:
\be
\tau_1 \to 0\,,\quad \tau_4 \to r_1 r_2\,, \quad
\tau_2 \to r_2^2+5\,r_2\,r_5 +5\,r_5^2\,, \quad
\tau_3 \to r_1 \left(r_2+5 r_5\right)+2 r_2 r_5+7 r_5^2\,,
\label{eq:Taus}
\ee
while the Calabi-Yau volume reduces to:
\be
 \vo\to r_1\tau_2+\tfrac 23 \,r_2^3 + 6\,r_2^2 \,r_5+102\,r_2\,r_5^2+88\,r_5^3\,.
\label{LimVol}
\ee
As we shall see in the next section, the volume of the K3 fibre is given by $\tau_2$
while the volume of the $\mbb P^1$ base of the fibration is controlled by the K\"ahler parameter $r_1$.
Therefore the anisotropic limit $r_1\gg \sqrt{\tau_2}$ corresponds to
$r_1\gg r_2\sim r_5$. In this new limit the expressions (\ref{eq:Taus}) and (\ref{LimVol}) further simplify to:
\be
\tau_1 \to 0\,,\quad \tau_4 \to r_1 r_2\,, \quad
\tau_2 \to r_2^2+5\,r_2\,r_5 +5\,r_5^2\,, \quad
\tau_3 \to r_1 \left(r_2+5 r_5\right)\,,\quad \vo\to r_1\tau_2. \nonumber
\label{TAus}
\ee
We can now invert the expression of the $r$'s in terms of the $\tau$'s obtaining:
\be
r_1 =\frac{\sqrt{\tau_3^2 + 3 \tau_3 \tau_4 + \tau_4^2}}{\sqrt{5 \tau_2}}\,, \quad
r_2 = \frac{\tau_4\sqrt{5 \tau_2}}{\sqrt{\tau_3^2 + 3 \tau_3 \tau_4 + \tau_4^2}}\,, \quad
r_5 =\frac{(\tau_3 - \tau_4)\sqrt{\tau_2} }{\sqrt{5} \sqrt{\tau_3^2 + 3 \tau_3 \tau_4 + \tau_4^2}}. \nonumber
\label{rs}
\ee
Hence, the Calabi-Yau volume in the limit $r_1\gg r_2\sim r_5\gg r_3\sim r_4>0$ takes the final form:
\be
\vo  = \sqrt{\frac {\tau_2}{5}} \left(\tau_3^2 + 3 \tau_3 \tau_4 + \tau_4^2\right)^{1/2}-\frac 13 \,\tau_1^{3/2}.
\ee

\subsubsection{Divisor analysis}

In this section we shall analyse the topology of the divisors of the Calabi-Yau three-fold
under consideration starting from the rigid divisors and then turning to the study of
the fibration structure.

By inspecting the weight matrix (\ref{eq:model2dP:weightm}), we notice that
this manifold has four rigid divisors: $D_1$, $D_3$, $D_4$ and $D_5$.
However only $D_1$ and $D_4$ fulfill the necessary condition~\eqref{poscurve} to be a del Pezzo divisor.
In fact, the topological indices of these surfaces indicate that they are a dP$_7$ and dP$_1$, respectively.
In order to make this more explicit, we proceed as above and project the parts of the fans that have
the respective ray as a subset along this ray. From the analysis of the so obtained fans,
together with the transformed hypersurfaces\footnote{Note again that the homogeneous coordinates
not intersecting the respective divisor are omitted.}, we get for the two surfaces:
\begin{align}
D_1:&\quad\begin{array}{|c|c|c|c||c|}\hline
z_5 & z_6 & z_7 & z_8 & D_\textmd{H}\big|_{D_1}\\\hline
1 & 2 & 1 & 1 & 4\\\hline\end{array}\quad\textmd{with}\quad  {\rm SR}\Big|_{D_1}=\{z_5\,z_6\,z_7\,z_8\}\,,
\quad\textmd{and}\\
D_4:&\quad\begin{array}{|c|c|c|c|c||c|}\hline
z_2 & z_3 & z_5 & z_7 & z_8 & D_\textmd{H}\big|_{D_4}\\\hline
0 & 3 & 0 & 1 & 1 & 3\\\hline
1 & 0 & 1 & 0 & 0 & 0 \\\hline\end{array}\quad\textmd{with}\quad  {\rm SR}\Big|_{D_4}=\{z_2\,z_5,\,z_3\,z_7\,z_8\}\,.
\end{align}
Hence $D_1$ is a del Pezzo seven surface while $D_4$ is a $\mbb P^1\times \mbb P^1$.

In order to find the K3 fibre, we search, by means of PALP, for all reflexive sections
of co-dimension one of the N-lattice polytope
which implies that the dual M-lattice polytope has a reflexive projection.
Given that the M-lattice polytope encodes the information about the monomials of the hypersurface equation,
the projection can be interpreted as the original equation with the homogeneous coordinates,
not on the section, set to some generic value.
The reflexivity then assures that the equation, in the homogeneous coordinates lying on the section,
is a Calabi-Yau two-fold, i.e.~a K3 surface. By using the theorem of~\cite{Oguiso}
we can then easily see whether this is also a fibration.

In order to make this more explicit in terms of toric language,
we need that the projection along the section of the polytope
maps the fan of the toric four-fold into a valid fan.
This toric homomorphism induces then the projection of the fibration.
The polytope under consideration has one section of this kind and the only points that do not lie on this plane
are the ones corresponding to $z_1$, $z_2$ and $z_5$.
From the projection along this plane we obtain the two vectors
of the $\mbb P^1$ where $z_1$ and $z_5$ are mapped to the same point.
Hence the homogeneous coordinates of the base $\mbb P^1$ are $\zeta_1=z_2$ and $\zeta_2=z_1\,z_5$.
In order to see that this projection is also a homomorphism between the fans,
we inspect the SR-ideals of the two toric varieties.
We have to check whether SR$_{\mbb P^1}=\{\zeta_1\,\zeta_2\}\simeq\{z_1\,z_2,\,z_2\,z_5\}$
is a subset of the SR-ideal~\eqref{eq:model2dP:sr-ideal} of the original toric four-fold.
Given that this is indeed the case, we have a globally defined projection
of the four-dimensional toric variety to $\mbb P^1$. The corresponding three-dimensional toric fibre:
\be
 \begin{array}{|c|c|c|c|c||c|}\hline
z_3 & z_4 & z_6 & z_7 & z_8 & D_\textmd{H}\big|_\textmd{fibre}\\\hline
1 & 0 & 2 & 1 & 1 & 5\\\hline
0 & 1 & 3 & 1 & 1 & 6 \\\hline\end{array}\quad\textmd{with}\quad  {\rm SR}\Big|_\textmd{fibre}=\{z_4 z_6,\,z_3 z_7 z_8\}\,,
\ee
degenerates at the point $\zeta_2=0$ into two toric divisors, $D_1$ and $D_5$, while
at $\zeta_1=0$ we have the generic fibre. Thus, together with the hypersurface equation,
we obtain a generic K3 fibre over this point. For this fibration we have also global sections.
Given that the three $\mbb P^1$'s turn out to be $\mc C_{3\,\mbb{P}^1}=D_3\cap D_4$,
the volume of the base is:
\be
\textmd{Vol}({\rm base})= \tfrac 13 \,\textmd{Vol}(\mc C_{3\,\mbb{P}^1})
=\tfrac 13\,\int_{ D_3\cap D_4} \imath^* J=\tfrac 13\,D_3\cdot D_4\cdot J=r_1=t_2-2t_4.
\ee
Let us now summarise the results of our divisor analysis:
\be
 \begin{array}{|c|c|c|c|}
\hline
\textmd{divisor classes} & D_1 & D_2\simeq D_1 + D_5 & D_4\\ \hline
{\rm variety}& \textmd{dP}_7 & \textmd{K3} & \mbb P^1\times \mbb P^1\\ \hline
{\rm volume} & 4\,(r_3+r_4)^2 & \begin{array}{c}
(r_2+r_3+3\,r_4+r_5)(r_2+5\,r_3+3\,r_4+5\,r_5)\\-(r_2+3\,r_4)(r_3+r_5) \end{array} & r_1\,(r_2+3\,r_4)\\ \hline
\end{array}\,.\nonumber
\ee

\subsubsection{Orientifolding}

We shall now compare the $\mbb Z_2$-involutions $z_6\leftrightarrow -z_6$ and $z_7\leftrightarrow -z_7$.
For the two cases, the generic hypersurface equation, expanded in the transformed coordinate, reduces to:
\bea
\label{eq:model1dP:cyhdegenerate}
 P_\textmd{CY}\big|_{z_6\leftrightarrow -z_6}&=&z_3\,z_6^2+z_4\,P_{(8,12,5,5)}=0\,,\quad\textmd{and}\\
 P_\textmd{CY}\big|_{z_7\leftrightarrow -z_7}&=&z_3\,P_{(8,12,4,6)}+z_8\,P_{(7,10,4,5)}+z_7^2\,(z_3\,P_{(6,8,2,4)}+z_8\,P_{(5,6,2,3)})\nonumber\\
&&+z_7^4\,(z_3\,P_{(4,4,0,2)}+z_8\,P_{(3,2,0,1)})\,,\nonumber
\eea
where the $P_{(i,j,k,l)}$ are generic polynomials, in the rest of the homogeneous coordinates,
of the indicated class. The fixed point sets of the toric four-fold turns out to be:
\bea
\{\textmd{Fixed}\}\big|_{z_6\leftrightarrow -z_6}&=&\{D_6,\,D_3\cdot D_4,\,D_4\cdot D_7\cdot D_8,
\,D_1\cdot D_5\cdot D_7\cdot D_8\}\quad\textmd{and}\nonumber\\
\{\textmd{Fixed}\}\big|_{z_7\leftrightarrow -z_7}&=&\{D_7,\,D_3\cdot D_8,\, D_1\cdot D_5\cdot D_8\}\,.
\eea
However, only $D_6$ and $D_3\cdot D_4$ in the first case, and $D_7$, $D_3\cdot D_8$,
and $D_1\cdot D_5\cdot D_8$ in the second case, intersect the respective hypersurface.
From~\eqref{eq:model1dP:cyhdegenerate}, we observe that $D_3\cdot D_4$ and $D_3\cdot D_8$
induce a second $O7$-plane on the corresponding Calabi-Yau hypersurfaces
in addition to $D_6$ and $D_7$.
Moreover, in the case of $z_7\leftrightarrow -z_7$, $D_1\cdot D_5\cdot D_8$ induces an $O3$-plane.

We finally stress that in the orientifold case nothing severely happens for the divisors that we considered above.
However, the toric divisor $D_3$ intersecting the hypersurface factorises into two parts in the orientifold case $z_7\leftrightarrow -z_7$,
as can be deduced from~\eqref{eq:model1dP:cyhdegenerate}.

\subsection{A K3 fibration with one del Pezzo divisor}
\label{sec:6.2}

The second non-simplicial example that we work out in detail is a K3 fibration
which admits only one toric del Pezzo divisor.
The fan of the toric four-fold is defined in terms of its weight matrix:
\be
\begin{array}{|c|c|c|c|c|c|c|c||c|}
\hline z_1 & z_2 & z_3 & z_4 & z_5 & z_6 & z_7 & z_8 & D_\textmd{H} \tabularnewline \hline \hline
0 & 0 & 1 & 0 & 0 & 1 & 1 & 3 & 6\tabularnewline\hline
1 & 2 & 0 & 0 & 1 & 1 & 0 & 5 & 10\tabularnewline\hline
1 & 1 & 1 & 0 & 0 & 1 & 0 & 4 & 8\tabularnewline\hline
0 & 0 & 0 & 1 & 0 & 0 & 1 & 2 & 4\tabularnewline\hline
\end{array}\label{eq:model1dP:weightm}\,,
\ee
and its Stanley-Reisner ideal:
\be
\label{eq:model1dP:sr-ideal}
{\rm SR}=\{z_1 z_2,\,z_3 z_6, \, z_1 z_3,\, z_5 z_6 z_7,
\, z_2 z_5 z_6,\, z_2 z_5 z_8,\, z_4 z_7 z_8,\, z_1 z_4 z_8\}\,.
\ee
With these data we can calculate the intersection structure on the Calabi-Yau hypersurface
in terms of the basis divisors $D_i=\{z_i=0\}$, $i=1,...,4$:
\be
 I_3= D_1^3+2\,D_2\,D_3\,D_4-2\,D_3^2\,D_4 -2\,D_2\,D_4^2+ 2\,D_4^3\,.
\ee
The Mori cone generators of the toric four-fold given by~\eqref{eq:model1dP:weightm} and \eqref{eq:model1dP:sr-ideal} are:
\be
\label{eq:model1dP:moricone}
\int_{\mc C_j} D_i=\left(\begin{array}{cccc}
0 & 1 & -1 & 0\\
1 & 0 & 0 & 2\\
-2 & 0 & 0 & -3\\
0 & -2 & 2 & 1\\
2 & 2 & 0 & -1
\end{array}\right):=\mc M_{ji}\,,
\ee
while the dual generators to~\eqref{eq:model1dP:moricone} look like:
\begin{eqnarray}
&&
\Gamma_1=D_2 + D_3\,,\quad
\Gamma_2= 4\,D_2 + 3\,D_3 + 2\,D_4-3\,D_1\,,\quad
\Gamma_3= 4\,D_2 + 4\,D_3 + 2\,D_4-3\,D_1 \,, \nonumber\\
&&
\Gamma_4= 5\,D_2 + 4\,D_3 + 2\,D_4-4 \,D_1\,,\quad\,
\Gamma_5= 5\,D_2 + 5\,D_3 + 2\,D_4-4 \,D_1\,.
\end{eqnarray}
The fact that we found five, instead of just four, $\Gamma$'s shows that
the K\"ahler cone is again non-simplicial. Expanding the K\"ahler form in terms of
the $\Gamma$'s as:
\be
 J=\sum_{j=1}^5 r_j \,\Gamma_j\quad\textmd{with}\quad r_j>0\,,
\ee
the divisor volumes turn out to be:
\bea
\label{Letaus}
\tau_1 &=& \tfrac 12 [3 r_2 + 3 r_3 + 4 (r_4 + r_5)]^2\,,\nonumber \\
\tau_2 &=& 4 (r_2 + r_3 + r_4 + r_5) [r_1 + 2 r_2 + 3 (r_3 + r_4) + 4 r_5]\,, \nonumber \\
\tau_3 &=& 4 (r_2 + r_4) (r_2 + r_3 + r_4 + r_5)\,, \\
\tau_4 &=& r_ 1 [r_ 1 + 4 r_ 2 + 4 r_ 3 + 6 (r_ 4 + r_ 5)] +
 r_ 2 (5 r_ 2 + 12 r_ 3 + 14 r_ 4 + 16 r_ 5) \nonumber \\
 &&+ 2 r_ 3[3 r_ 3 + 8 (r_ 4 +   r_ 5)] + 2 r_ 4 ( 5 r_ 4 + 11 r_ 5) +
 11 r_ 5^2\,, \nonumber
\eea
whereas the volume of the Calabi-Yau manifold becomes:
\bea
 \vo&=&\frac{1}{6} \,(r_2 + r_3) (73\, r_2^2 + 170\, r_2\, r_3 + 85\, r_3^2) +
 2\, (22\, r_2^2 + 48\, r_2\, r_3 + 25\, r_3^2)\, r_4 \nonumber\\
&+& 4\, (13\, r_2 + 14\, r_3)\, r_4^2
+ 20\, r_4^3 +
 2\, [24\, r_2^2 + 25\, r_3^2 + 58\, r_3\, r_4 + 32\, r_4^2 \\
 &+&    r_2\, (50\, r_3 + 56\, r_4)]\, r_5 +
 2\, [29\, (r_2 + r_3) + 33\, r_4]\, r_5^2 + 22\, r_5^3 +
 2 \,r_1^2\, (r_2 + r_3 + r_4 + r_5) \nonumber\\
 &+& 4\, r_1\, (r_2 + r_3 + r_4 + r_5) [3\, r_2 + 3\, r_3 +
    4\, (r_4 + r_5)]+\tfrac 43 (r_2 + r_3 + r_4 + r_5)^3\,.\nonumber
\label{IlNuovoVolume}
\eea
As we pointed out in the previous section, in order to write down the
four-dimensional effective field theory we need to expand the K\"ahler form
in terms of the initial basis $D_i$, $i=1,...,4$ as:
\be
J=\sum_{i=1}^4 t_i\,D_i\,,\quad\textmd{with}\quad \sum_{i=1}^4\mc{M}_{ji} t_i>0\,.
\ee
The volume of the basis divisor and the Calabi-Yau three-fold as a function of the four linear independent $t$'s looks like:
\bea
\label{TAUS}
\tau_1 &=& \tfrac 12 \,t_1^2\,,\qquad \tau_2 = t_4 (2 t_3 - t_4)\,, \\
\tau_3 &=& 2 t_4 (t_2 - t_3)\,,\quad \tau_4 = \tfrac 32\,  t_4^2 -t_3^2+2 t_2 (t_3 - t_4)\,, \nonumber
\eea
and:
\bea
\label{VOLUME}
\vo&=&\frac 16 \left[3 t_4 (4 t_2 t_3 - 2 t_3^2 - 2 t_2 t_4 + t_4^2)+t_1^3\right]\nonumber \\
&=& t_2 \tau_2 + \frac {t_4}{2} \,(t_4^2 - 2 t_3^2)-\frac{\sqrt{2}}{3}\,\tau_1^{3/2}.
\eea
Similarly to the previous non-simplicial example, we have not been able
to write the volume just in terms of K\"ahler moduli $\tau_i$. Thus
we did not manage to diagonalise the volume even if the modulus $\tau_1$ is
manifestly diagonal. As we shall see in the next section,
the divisor $D_1$ is a del Pezzo eight surface.

Let us now show that, by taking the limit of small $\tau_1$,
the volume simplifies considerably, and so it can easily be written just in terms of the $\tau$'s.
As can be seen from (\ref{Letaus}), shrinking $\tau_1$ corresponds to
taking $r_1\gg r_2\sim r_3\sim r_4\sim r_5 \sim r>0$. In this limit, the
divisor volumes become:
\be
\tau_1 \sim 98 \, r^2\,,\quad \tau_2 \sim 16\, r\, r_1\,, \quad
\tau_3 \sim 32 \,r^2\,, \quad
\tau_4 \sim r_1^2\,\quad\Rightarrow\quad \tau_4\gg \tau_2 \gg \tau_1 \gtrsim \tau_3\,,
\label{Taus}
\ee
and so we notice that also the four-cycle $\tau_3$ is forced to shrink.
On the other hand, the two-cycle parameters take the form:
\bea
t_1 &=& -(3 r_2 + 3 r_3 + 4  r_4 + 4 r_5) \sim - 14\, r\,,\nonumber \\
t_2 &=& r_1 + 4 r_2 + 4 r_3 + 5 r_4 + 5 r_5 \sim  r_1 + 18\, r\,, \nonumber \\
t_3 &=& r_1 + 3  r_2+ 4 r_3 + 4 r_4 + 5 r_5 \sim r_1 + 16 \,r\,, \\
t_4 &=& 2 r_2 + 2 r_3 + 2 r_4 + 2 r_5\sim 8 \,r\,,\quad\Rightarrow\quad t_2 \gtrsim t_3 \gg  |t_1| \gtrsim t_4. \nonumber
\eea
Therefore the relations (\ref{TAUS}) simplify to:
\be
\tau_1 \sim \tfrac 12 \,t_1^2\,,\qquad \tau_2 \sim 2 t_3 t_4\,, \quad
\tau_3 \sim 2 t_4 (t_2 - t_3)\,,\quad \tau_4 \sim  t_3 (2 \,t_2-t_3) \,, \nonumber
\ee
while the Calabi-Yau volume reduces to:
\be
\label{LIMVOLUME}
\vo= t_3 \, t_4 ( 2 t_2 -  t_3 )+\frac 16\, t_1^3
= \frac {( 2 t_2 -  t_3 )}{2} \,\tau_2  -\frac{\sqrt{2}}{3}\,\tau_1^{3/2}
= \frac 12\,\sqrt{\tau_4 \tau_2 \left(\tau_2+ 2 \tau_3\right)} -\frac{\sqrt{2}}{3}\,\tau_1^{3/2},
\ee
where, as we shall see in the next section, the volume of the K3 fibre is given by $\tau_2$.
We conclude this section by showing that this limit corresponds to the anisotropic limit of
\cite{Cicoli:2011yh,Cicoli:2011yy} since:
\be
\frac {( 2 t_2 -  t_3 )}{2} \sim \frac {r_1}{2} \gg \sqrt{\tau_2}\sim 4 \sqrt{r \,r_1}.
\ee

\subsubsection{Divisor analysis}

We now turn to the study of the topology of the divisors of the Calabi-Yau
three-fold under consideration.

By looking at the weight matrix (\ref{eq:model1dP:weightm}),
we realise that the rigid divisors of the Calabi-Yau three-fold
are $D_1$, $D_3$, $D_4$, $D_5$ and $D_7$. However, only $D_1$ is a del Pezzo.
Its topological numbers, $\chi_h=1$ and $\chi=11$, indicate
that this divisor is a dP$_8$.
In order to check this, we project the relevant part of the fan along $z_1$ obtaining:
\be
D_1:\quad\begin{array}{|c|c|c|c|c||c|}\hline
z_4 & z_5 & z_6 & z_7 & z_8 & D_\textmd{H}\big|_{D_1}\\\hline
0 & 1 & 1 & 2 & 3 & 6\\\hline
1 & 0 & 0 & 1 & 2 & 4 \\\hline\end{array}\quad\textmd{with}
\quad  {\rm SR}\Big|_{D_1}=\{z_5 z_6 z_7,\,z_4 z_7 z_8\} \,,
\ee
which shows explicitly that $D_1$ is a del Pezzo eight surface.

In order to find the K3 fibration we look for the reflexive sections
of the N-lattice polytope encoded in~\eqref{eq:model1dP:weightm}
finding that the polytope at hands admits just one section of this type.
The vectors corresponding to the homogeneous coordinates $z_1$, $z_2$ and $z_5$ do not lie on this plane.
The projection along the plane maps $z_2$ to $\zeta_1$ and $z_1\,z_5$ to $\zeta_2$,
where $\zeta_1$ and $\zeta_2$ are the homogeneous coordinates of the $\mbb P^1$.
In this case the projection does not define a toric homomorphism
because SR$_{\mbb P^1}=\{\zeta_1\,\zeta_2\}\simeq\{z_1\,z_2,\,z_2\,z_5\}$ is not a subset of~\eqref{eq:model1dP:sr-ideal}.
However, the Calabi-Yau hypersurface equation does not go through the points $z_2=z_5=0$.
Hence we may remove all the cones from the fan that contain the two-cone $\langle v_2 v_5\rangle$.
Even if the toric four-fold of the new fan is not compact anymore, the hypersurface is the same.
For the new toric four-fold, $z_2\,z_5$ is part of the SR-ideal,
and so we obtain a valid projection. This is in agreement with the theorem of~\cite{Oguiso}.
Over the point $\zeta_1=z_2=0$, we get a generic K3 fibre and at $\zeta_2=0$
the fibre becomes the sum of $D_1$ and $D_5$.
We finally point out that for this fibration
it is not possible to find a section that embeds the base into the Calabi-Yau.

Let us now summarise the results of our divisor analysis:
\be
 \begin{array}{|c|c|c|c|}
\hline
\textmd{divisor classes} & D_1 & D_2\simeq D_1 + D_5 \\ \hline
{\rm variety} & \textmd{dP}_8 & \textmd{K3} \\ \hline
{\rm volume} & \tfrac{1}{2}[3 r_2 + 3 r_3 + 4 (r_4 + r_5)]^2
& 4\, (r_2 + r_3 + r_4 + r_5) [r_1 + 2 r_2 + 3 (r_3 + r_4) + 4 r_5] \\ \hline
\end{array}\,.\nonumber
\ee

\subsubsection{Orientifolding}

We again briefly study the orientifold case by comparing the two different
involutions $z_6\leftrightarrow -z_6$ and $z_8\leftrightarrow -z_8$.
For the fixed point sets of the ambient space, we obtain:
\bea
\{\textmd{Fixed}\}\big|_{z_6\leftrightarrow -z_6}&=&\{D_6,
\,D_3\cdot D_4\cdot D_8,
\,D_1\cdot D_5\cdot D_8,
\,D_2\cdot D_4\cdot D_5,
\,D_3\cdot D_7\cdot D_8,\nonumber\\&&
\,D_3\cdot D_4\cdot D_7,
\,D_2\cdot D_5\cdot D_7,
\,D_1\cdot D_4\cdot D_5\cdot D_7
\}\quad\textmd{and}\nonumber\\
\{\textmd{Fixed}\}\big|_{z_8\leftrightarrow -z_8}&=&\{D_8,\,D_4\cdot D_7,\, D_1\cdot D_5\cdot D_6,\, D_2\cdot D_3\cdot D_5\}\,.
\eea
However only $D_6$, $D_3\cdot D_4\cdot D_8$, $D_1\cdot D_5\cdot D_8$ and $D_3\cdot D_7\cdot D_8$ in the first case,
and $D_8$ and $D_1\cdot D_5\cdot D_6$ in the second case, intersect the respective invariant hypersurface.
Thus we get in both cases at least one $O3$-plane in addition to the respective $O7$-plane.
Regarding the Calabi-Yau hypersurface, also the K3 fibre and the del Pezzo eight surface become less generic
even if, together with all other toric divisors, they do not degenerate.

We finally have a quick look at the F-theory up-lift in the case of an orientifold with $z_8\leftrightarrow -z_8$ involution.
By looking at the weight matrix~\eqref{eq:model1dP:weightm}, we realise that this Calabi-Yau three-fold has a structure similar
to the one analysed in~\cite{Collinucci:2008pf}.
The $\mbb Z_2$-quotient of the Calabi-Yau is the toric three-fold of~\eqref{eq:model1dP:weightm} with $z_8$ omitted.
Furthermore, the Calabi-Yau four-fold is given by a hypersurface in:
\begin{equation}
\begin{array}{|c|c|c||c|c|c|c|c|c|c||c|}
\hline
\phantom{.}y\phantom{.} & \phantom{.}x\phantom{.} & \phantom{.}z\phantom{.}
& z_1 & z_2 & z_3 & z_4 & z_5 & z_6 & z_7 & \sum_{Y_4}\tabularnewline
\hline\hline
 3 & 2 & 1 & 0 & 0 & 0 & 0 & 0 & 0 & 0 & 6 \\ \hline\hline
 9 & 6 & 0 & 0 & 0 & 1 & 0 & 0 & 1 & 1 & 6 \\ \hline
 15 & 10 & 0 & 1 & 2 & 0 & 0 & 1 & 1 & 0 & 10 \\ \hline
 12 & 8 & 0 & 1 & 1 & 1 & 0 & 0 & 1 & 0 & 8 \\ \hline
 6 & 4 & 0 & 0 & 0 & 0 & 1 & 0 & 0 & 1 & 4 \\ \hline
\end{array}\,.
\end{equation}
The Hodge number $h^{1,1}(Y_4)$ increased by one compared to $h_+^{1,1}(Y_3)=h^{1,1}(Y_3)=4$ of the original three-fold.
The Euler characteristic of $Y_4$ is 13416 which is
in agreement with the $D3$-tadpole of a type IIB orientifold with one Whitney brane~\cite{Collinucci:2008pf} that cancels the $D7$-tadpole.
We stress that in this evaluation we also included the contribution to the tadpole coming from $O3$-planes
finding that the stringy Hodge number calculation of~\cite{Batyrev:1995ca} does indeed take the contribution of $O3$-planes into account.
This is in contrast to~\cite{Collinucci:2008zs} where the Euler number calculation
matches the tadpole only without the $O3$-planes.\footnote{We are grateful to Sven Krause
for doing the four-fold calculations and observing the mismatch.}

\section{Conclusions}

In this paper we used the powerful tools of toric geometry to construct
several explicit examples of K3 fibrations which can be used
in the study of Calabi-Yau string compactifications.

Historically this kind of manifolds received a lot of attention due to the crucial
r\^ole that they play in the investigation of string dualities \cite{Klemm:1995tj,Kachru:1995wm,Sen:1995ff,Aspinwall:1995vk,Aldazabal:1995yw,Hunt:1995sy,Aspinwall:1996mw,Gomez:1996xi}.
Therefore many explicit examples of K3-fibred Calabi-Yau
three-folds have been built as hypersurfaces embedded
either in complex weighted projective spaces \cite{Candelas:1993dm,Klemm:1995tj,Hosono}
or, more generally, in toric varieties \cite{Avram:1996pj}.
The present work extends these constructions to include the interesting case of
K3 fibrations with the presence of additional del Pezzo divisors.

The main motivation for these constructions is the
relatively recent discovery of new LARGE Volume type IIB vacua \cite{Balasubramanian:2005zx}
for K3-fibred Calabi-Yau three-folds \cite{Cicoli:2008va} which turn out to be very promising
for several applications to cosmology \cite{Cicoli:2008gp,Burgess:2010bz,Cicoli:2010ha,Anguelova:2009ht}
and particle phenomenology \cite{Cicoli:2011yh,Cicoli:2011yy}.
As derived in \cite{Cicoli:2008va}, two conditions for the existence
of such minima at exponentially large value of the internal volume
are a negative Euler number, i.e. $h^{1,2} > h^{1,1} > 1$, and the
presence of at least one local blow-up mode
which is needed to support the non-perturbative effects that lead to moduli
stabilisation by their interplay with $\alpha'$ and $g_s$ corrections.
However, before this paper, no explicit examples of K3 fibrations with additional blow-up modes were known in the literature.

Given that the final goal is not just to achieve moduli stabilisation
but also to realise a chiral MSSM or GUT-like local construction with
intersecting $D$-branes wrapped around internal divisors, we looked
also for a second rigid divisor which could support chiral matter.

However, before performing our Calabi-Yau search,
we gave a detailed description of the procedure to diagonalise the volume form
and then we discussed briefly the interplay between a diagonal structure of the Calabi-Yau volume
and the supersymmetric requirement of having vanishing Fayet-Iliopoulos terms.
This discussion allowed us to classify the blow-up cycles in three different types:
\begin{enumerate}
\item[$i)$] `Diagonal' del Pezzo's (with only their triple self intersection number
non-vanishing, implying that they appear in the volume form in a completely
diagonal manner) which are natural candidates for supporting the non-perturbative effects of the LVS
or for leading to quiver GUT-like constructions due to their shrinking induced by the $D$-terms.

\item[$ii)$] `Non-diagonal' del Pezzo's (with non-zero intersections with different divisors in any basis,
implying that they do not appear in the Calabi-Yau volume in a completely diagonal way)
which are suitable to support GUT constructions with magnetised branes wrapping cycles stabilised in the geometric
regime since they are rigid and, above all, in this `non-diagonal' case,
the vanishing $D$-term condition might not force the shrinking of any cycle.

\item[$iii)$] Rigid but not del Pezzo divisors (not contractible to a point) which,
similarly to the `non-diagonal' del Pezzo divisors, cannot be used
to support the non-perturbative effects in the LVS, but are natural candidates
for GUT model building in the geometric regime.
\end{enumerate}
After this classification of blow-up cycles, we went through the
existing list of four dimensional reflexive lattice polytopes~\cite{Kreuzer:2000xy,cydata}
searching for K3 fibrations with four K\"ahler moduli where at least one of them
is a `diagonal' del Pezzo divisor. The final result of our search consists of
158 lattice polytopes corresponding to toric ambient varieties which admit a
Calabi-Yau hypersurface with the required features.
We then performed a detailed analysis of the topology of
some illustrative examples derived from both simplicial and non-simplicial polytopes.
Interestingly enough, we found the presence of `non-diagonal' rigid divisors
only in the case of non-simplicial polytopes.

We point out that this paper represents just a first step towards the
final goal of bringing together the existing scenarios
of moduli stabilisation and chiral $D$-brane model building.
The next step would be to use these compact Calabi-Yau
three-folds to try to achieve full closed string moduli stabilisation
in a globally consistent model
(tadpole and Freed-Witten anomaly free)
with an explicit brane set-up and a particular choice of world-volume fluxes
that lead to the local presence of chirality.
We will discuss these matters in more detail in a forthcoming companion article \cite{New}.

We finally stress that the main motivation behind this paper
has been to provide a more rigorous mathematical foundation to
the interesting LVS. However the
several explicit realisations of K3-fibrations
with additional del Pezzo divisors that we presented in this paper
could be very useful also for other completely different
model building scenarios in string compactifications.

\section*{Acknowledgements}

We would like to thank Sven Krause, Sven Krippendorf, Fernando Quevedo, Roberto Valandro,
Nils-Ole Walliser, and Timo Weigand for useful discussions
and especially Andr\'es Collinucci for initial collaboration on this project.
We thank the Abdus Salam International Centre for Theoretical Physics
for hospitality during the development of part of this work.
The work of CM was supported by the DFG through TRR33 ``The Dark Universe''.

\appendix

\section{Two K3 fibrations with an F-theory uplift}
\label{appA}

As a result of our search, 58 of the 158 K3-fibred Calabi-Yau manifolds
with at least one `diagonal' del Pezzo
divisor that we found, feature a F-theory uplift
in terms of an elliptically fibred Calabi-Yau four-fold.
In this appendix we shall present two simplicial examples
of these K3 fibrations with such an uplift.

\subsection{$\mbb P^4_{[1,1,1,1,4]}(8)$ with three blow-ups}
\label{sec:A.1}

In order to define the fan of the toric four-fold, we give its weight matrix:
\be
\begin{array}{|c|c|c|c|c|c|c|c||c|}
\hline z_1 & z_2 & z_3 & z_4 & z_5 & z_6 & z_7 & z_8 & D_\textmd{H} \tabularnewline \hline \hline
1 & 0 & 1 & 1 & 1 & 0 & 0 & 4 & 8\tabularnewline\hline
0 & 0 & 0 & 1 & 0 & 1 & 0 & 2 & 4\tabularnewline\hline
0 & 0 & 1 & 0 & 0 & 0 & 1 & 2 & 4\tabularnewline\hline
1 & 1 & 0 & 0 & 0 & 0 & 0 & 2 & 4\tabularnewline\hline
\end{array}\label{eq:modelP4dP:weightm}\,,
\ee
and its SR-ideal:
\be
\label{eq:modelP4dP:sr-ideal}
{\rm SR}=\{z_6\,z_7,\,z_2\,z_6,\,z_1\,z_2,\,z_4\,z_5,\,z_4\,z_6,\,z_3\,z_7\,z_8,\,z_1\,z_3\,z_5\,z_8\}.
\ee
The three blow-up divisors are given by $D_2$, $D_6$ and $D_7$.

The intersection structure on the Calabi-Yau hypersurface in terms of the basis divisors $D_i=\{z_i=0\}$, $i=1,...,4$ is:
\be
 I_3= -2\,D_1^3-2\,D_3^3-2\,D_2^2\,D_3+2\,D_3^2\,D_2+2\,D_1\,D_3\,D_4+2\,D_2\,D_3\,D_4\,,
\ee
while the generators of the Mori cone of the toric four-fold given by~\eqref{eq:modelP4dP:weightm} and \eqref{eq:modelP4dP:sr-ideal},
turn out to be:
\be
\label{eq:modelP4dP:moricone}
\int_{\mc C_j}D_i=\left(\begin{array}{cccc}
1 &  0 &  1 &  0 \\
0 &  -1 &  0 &  1 \\
-1 &  0 &  0 &  0 \\
1 &  1 &  -1 &  0 \\
\end{array}\right)\,.
\ee
We shall now expand the K\"ahler form in terms of
the generators $\Gamma_i$ of the K\"ahler cone as:
\be
J=\sum_{i=1}^4 r_i\,\Gamma_i \quad\textmd{with}\quad r_i>0,
\ee
where the generators dual to~\eqref{eq:modelP4dP:moricone}, look like:
\be
\Gamma_1= 2\,D_2 + D_3 + 2\,D_4-D_1,\quad
\Gamma_2=D_4,\quad
\Gamma_3=D_2 + D_4,\quad
\Gamma_4=D_2 + D_3 + D_4.
\ee
The volumes of the basis divisors are given by:
\bea
\tau_1 &=& 3 r_1^2 + 2 r_4 (r_2 + r_3 + r_4) + 2 r_1 (r_2 + r_3 + 3 r_4)\,,\quad
\tau_2 = (r_1 + r_4) (r_1 + 2 r_2 + r_4)\,,\nonumber\\
\tau_3 &=& (r_1 + r_3) (3 r_1 + 2 r_2 + r_3) + 2 r_4 (3 r_1 + r_2 + 2 r_3) + 2 r_4^2\,,\,\,
\tau_4 = 2 (r_1 + r_4) (r_1 + r_3 + r_4)\nonumber\,,
\eea
while the volume of the Calabi-Yau manifold reads:
\bea
 \vo&=& r_1 (r_1 + r_3) [2 (r_1 + r_2) + r_3] + (6 r_1^2 + 4 r_1 r_2 + 6 r_1 r_3 +
    2 r_2 r_3 + r_3^2) r_4 \nonumber \\
    &+& (6 r_1 + 2 r_2 + 3 r_3) r_4^2 + \tfrac 53\, r_4^3.
\eea
Following the algorithm presented in section~\ref{sec:diagonalize}, we find the diagonal basis:
\bea
D_a & := &  D_4 \,, \qquad
D_b  :=  2\,D_2 + D_4\,,\nonumber \\
D_c & := & 2\,(D_2 + D_3-D_1) + D_4 \,,\quad
D_d  :=  D_2+ D_4-D_1= D_6\,,
\eea
where the intersection polynomial reduces to:
\be
  I_3=8\,D_a\,D_b\,D_c+2\,D_d^3\,.
\ee
The Calabi-Yau volume simplifies to:
\be
\vo  = t_a \tau_a - \beta\, \tau_d^{3/2}=\alpha \sqrt{\tau_a\, \tau_b\, \tau_c} - \beta\, \tau_d^{3/2},
\ee
with $\alpha= 1/\left(2\sqrt{2}\right)$ and $\beta = 1/3$.
As we shall see in the next section,
$\tau_a=\tau_4$ gives the volume of the K3 fibre while $\tau_d=\tau_6$ is
the volume of a dP$_7$ diagonal divisor.

\subsubsection{Divisor analysis}

Let us now study the topology of the divisors of the Calabi-Yau three-fold under consideration.
The rigid divisors are $D_2$, $D_5$, $D_6$ and $D_7$.
However only $D_6$ is a del Pezzo since its topological numbers,
$\chi_h=1$ and $\chi=10$, indicate that this divisor is a dP$_7$.
In order to check this, we project the relevant part of the fan along $z_6$ obtaining:
\be
D_6:\quad\begin{array}{|c|c|c|c||c|}\hline
z_5 & z_1 & z_3 & z_8 & D_\textmd{H}\big|_{D_6}\\\hline
1 & 1 & 1 & 2 & 4\\\hline\end{array}\quad\textmd{with}\quad  {\rm SR}\Big|_{D_6}=\{z_1\,z_3\,z_5\,z_8\}\,,
\ee
which shows explicitly that $D_6$ is a del Pezzo seven surface.

In order to find the K3 fibration we look for the reflexive sections of the N-lattice polytope encoded in~\eqref{eq:modelP4dP:weightm}
which admits three sections of this kind even if only one is compatible with the projection.
The vectors corresponding to the homogeneous coordinates $z_4$, $z_5$ and $z_6$ do not lie on this plane.
The projection along the plane maps $z_4$ to $\zeta_1$ and $z_5\,z_6$ to $\zeta_2$,
where $\zeta_1$ and $\zeta_2$ are the homogeneous coordinates of the $\mbb P^1$.
This projection does indeed define a toric homomorphism because SR$_{\mbb P^1}=\{\zeta_1\,\zeta_2\}\simeq\{z_4\,z_5,\,z_4\,z_6\}$
is a subset of~\eqref{eq:modelP4dP:sr-ideal}.
Over the point $\zeta_1=z_4=0$, we get a generic K3 fibre and at $\zeta_2=0$ the fibre
becomes the sum of $D_5$ and $D_6$. We finally stress that,
for this fibration, it is not possible to find a section that embeds the base into the Calabi-Yau three-fold.

Let us summarise the results of our divisor analysis:
\be
 \begin{array}{|c|c|c|c|}
\hline
\textmd{divisor classes} & D_4\simeq D_5 + D_6   & D_6 \\ \hline
{\rm variety} &\textmd{K3} & \textmd{dP}_7 \\ \hline
{\rm volume} & 2 (r_1 + r_4) (r_1 + r_3 + r_4) & r_4^2 \\ \hline
\end{array}\,.\nonumber
\ee

\subsubsection{Orientifolding}

Let us now briefly discuss the case of the Calabi-Yau orientifold with
$z_8\leftrightarrow -z_8$ involution. The fixed point set of the ambient space is given by:
\bea
\{\textmd{Fixed}\}\big|_{z_8\leftrightarrow -z_8}&=&\{D_8,\,D_3\cdot D_7,\, D_1\cdot D_3 \cdot D_5 \cdot D_6\}\,.
\eea
However, only $D_8$ intersects the invariant hypersurface, and so we get only an $O7$-plane.

Regarding the Calabi-Yau hypersurface, also the K3 fibre and the del Pezzo seven surface
become less generic in the orientifold case even if, together with all the other toric divisors,
they do not degenerate.

For the F-theory uplift, the stringy Euler number of the four-fold is 14688 which is in agreement with a single
Whitney brane configuration.

\subsection{Another example with one del Pezzo divisor}
\label{sec:A.2}

The second example that we work out in detail in this appendix, is a K3 fibration
which admits only one toric del Pezzo divisor.
The fan of the toric four-fold is defined in terms of its weight matrix:
\be
\begin{array}{|c|c|c|c|c|c|c|c||c|}
\hline z_1 & z_2 & z_3 & z_4 & z_5 & z_6 & z_7 & z_8 & D_\textmd{H} \tabularnewline \hline \hline
0 & 1 & 0 & 0 & 2 & 2 & 6 & 1 & 12\tabularnewline\hline
0 & 0 & 0 & 1 & 1 & 1 & 3 & 0 & 6\tabularnewline\hline
0 & 0 & 1 & 0 & 0 & 1 & 2 & 0 & 4\tabularnewline\hline
1 & 1 & 0 & 0 & 1 & 0 & 3 & 0 & 6\tabularnewline\hline
\end{array}\label{eq:model2nddP8:weightm}\,,
\ee
and its SR-ideal:
\be
\label{eq:model2nddP8:sr-ideal}
{\rm SR}=\{z_1\, z_2,\,z_2\, z_8,\,z_1\, z_3,\,z_1\, z_4,\,z_4\, z_5,\,z_3\, z_6\, z_7,\,z_5\, z_6\, z_7\, z_8\}.
\ee
With these data we can calculate the intersection structure on the Calabi-Yau hypersurface
in terms of the basis divisors $D_i=\{z_i=0\}$, $i=1,...,4$:
\be
 I_3= D_1^3-2\, D_2 \, D_3^2+4\,D_3^3+2\,D_2 \,D_3\,D_4-4 \,D_3\,D_4^2\,.
\ee
The Mori cone generators of the toric four-fold given by~\eqref{eq:model2nddP8:weightm}
and \eqref{eq:model2nddP8:sr-ideal} are:
\be
\label{eq:model2nddP8:moricone}
\int_{\mc C_j}D_i=\left(\begin{array}{cccc}
-1 &  0 &  0 &  0 \\
0 &  1 &  0 &  -2 \\
0 &  0 &  -3 &  2 \\
2 &  0 &  3 &  0 \\
\end{array}\right):=\mc{M}_{ji}\,,
\ee
while the dual generators to~\eqref{eq:model1dP:moricone} look like:
\be
\Gamma_1=6\, D_2 - 3 \,D_1 + 2 \,D_3 + 3\,D_4,\quad
\Gamma_2=D_2 ,\quad
\Gamma_3=2\,D_2 + D_4,\quad
\Gamma_4=6\,D_2 + 2\,D_3 + 3\,D_4.
\ee
Expanding the K\"ahler form in terms of
the $\Gamma$'s as:
\be
 J=\sum_{i=1}^4 r_i \,\Gamma_i,\quad\textmd{with}\quad r_i>0\,,
\ee
the divisor volumes turn out to be:
\bea
\label{taui}
\tau_1 &=& \tfrac 92 \, r_1^2\,,\qquad
\tau_2 = 4 (r_1 + r_4) (2 \,r_1 + r_3 + 2\, r_4)\,, \\
\tau_3 &=& 2 (r_1 + r_3 + r_4) (r_1 + r_2 + r_3 + r_4)\,, \quad
\tau_4 = 4 r_2 (r_1 + r_4)\,, \nonumber
\eea
whereas the volume of the Calabi-Yau manifold becomes:
\bea
 \vo&=&
\tfrac{77}{6}\,r_1^3 + 4\,r_3\,(r_2 + r_3)\,r_4 + 8\,(r_2 + 2\,r_3)\,r_4^2
+ \tfrac{52}{3}\,r_4^3 + 4\,r_1^2\,(2\,r_2 + 4\,r_3 + 13\,r_4)\nonumber\\
&+& 4\,r_1\,[r_3\,(r_2 + r_3) + 4\,(r_2 + 2\,r_3)\,r_4 + 13\,r_4^2].
\eea
These expressions simplify considerably if we expand the K\"ahler form
in terms of the initial basis $D_i$, $i=1,...,4$ as:
\be
J=\sum_{i=1}^4 t_i\,D_i\,,\quad\textmd{with}\quad \sum_{i=1}^4\mc{M}_{ji} t_i>0\,,
\ee
since they become:
\bea
\label{TAUi}
\tau_1 &=& \tfrac 12 \,t_1^2\,,\qquad \tau_2 = t_3 ( 2 \,t_4-t_3)\,, \\
\tau_3 &=& 2 (t_3 - t_4) (t_3 + t_4-t_2)\,,\quad \tau_4 = 2 t_3 (t_2 - 2 t_4)\,, \nonumber
\eea
and:
\bea
\label{VOLUMe}
\vo&=&\tfrac 13 \,t_3 ( 2 \,t_3^2 -3 \,t_2 \,t_3 + 6 \,t_2 \,t_4 - 6 \,t_4^2)+\tfrac 16 t_1^3 \nonumber \\
&=& (t_2-2\,t_4) \tau_2 + \frac 23 \,t_3(t_3^2 - 3 \,t_3 \,t_4 + 3 \,t_4^2)-\frac{\sqrt{2}}{3}\,\tau_1^{3/2}.
\eea
Similarly to the previous example, we have not been able
to write the volume just in terms of K\"ahler moduli $\tau_i$. Thus
we did not manage to diagonalise the volume even if the modulus $\tau_1$ is
manifestly diagonal. As we shall see in the next section,
the divisor $D_1$ is a del Pezzo eight surface.

Let us now show that we can indeed write down the volume form just in terms of the $\tau$'s
in the anisotropic large volume limit.
We start by taking the limit of small $\tau_1$. As can be seen from (\ref{taui}),
this is equivalent to sending $r_1\to 0$. In this limit, the
divisor volumes become:
\be
\label{TAUs}
\tau_1 \to 0\,,\quad \tau_2 \to 4 \, r_4 ( r_3 + 2\, r_4)\,, \quad
\tau_3 \to 2 (r_3 + r_4) (r_2 + r_3 + r_4)\,, \quad
\tau_4 \to 4 \,r_2 \, r_4\,,
\ee
while the Calabi-Yau volume reduces to:
\be
 \vo\to r_2 \tau_2+4\,r_3\,r_4(r_3+ 4 \,r_4)+ \tfrac{52}{3}\,r_4^3.
\label{LImVol}
\ee
As we shall see in the next section, the volume of the K3 fibre is given by $\tau_2$
while the volume of the $\mbb P^1$ base of the fibration is controlled by the K\"ahler parameter $r_2=t_2-2\,t_4$.
Therefore the anisotropic limit $r_2\gg \sqrt{\tau_2}$ corresponds to
$r_2\gg r_3\sim r_4$. In this new limit the expressions (\ref{TAUs}) and (\ref{LImVol}) further simplify to:
\be
\tau_1 \to 0\,,\quad \tau_2 \to 4 \, r_4 ( r_3 + 2\, r_4)\,, \quad
\tau_3 \to 2\,r_2\, (r_3 + r_4)\,, \quad
\tau_4 \to 4 \,r_2 \, r_4\,,\quad \vo\to r_2\tau_2. \nonumber
\label{TAui}
\ee
We can now invert the expression of the $r$'s in terms of the $\tau$'s obtaining:
\be
r_2 =\frac{\sqrt{\tau_4 \left(2 \tau_3 + \tau_4\right)}}{2 \sqrt{\tau_2}}\,, \quad
r_3 = \frac{\sqrt{\tau_2} \left(2 \tau_3 - \tau_4\right)}{2 \sqrt{\tau_4\left(2 \tau_3 + \tau_4\right)}}\,, \quad
r_4 =\frac{\sqrt{\tau_2\,\tau_4}}{2\sqrt{2\tau_3 + \tau_4}}, \nonumber
\label{Erres}
\ee
and so the Calabi-Yau volume in the limit $r_2\gg r_3\sim r_4\gg r_1>0$ takes the final form:
\be
\vo  = \frac 12 \sqrt{\tau_2 \tau_4 \left(2 \tau_3 + \tau_4\right)}-\frac {\sqrt{2}}{3} \,\tau_1^{3/2}.
\ee

\subsubsection{Divisor analysis}

In this section we shall analyse the topology of the divisors of the Calabi-Yau three-fold under consideration.
The rigid divisors are $D_1$, $D_3$, $D_4$ and $D_8$.
However only $D_1$ is a del Pezzo since its topological numbers, $\chi_h=1$ and $\chi=11$,
indicate that this divisor is a dP$_8$.
In order to check this, we project the relevant part of the fan along $z_1$ obtaining:
\be
D_1:\quad\begin{array}{|c|c|c|c||c|}\hline
z_5 & z_6 & z_7 & z_8 & D_\textmd{H}\big|_{D_1}\\\hline
1 & 2 & 3 & 1 & 6\\\hline\end{array}\quad\textmd{with}\quad  {\rm SR}\Big|_{D_1}=\{z_5\, z_6\, z_7\, z_8\}\,,
\ee
which shows explicitly that $D_1$ is a del Pezzo eight surface.

In order to find the K3 fibration we look for the reflexive sections
of the N-lattice polytope encoded in~\eqref{eq:model2nddP8:weightm}
which admits two sections of this kind even if only one is compatible with the projection.
The vectors corresponding to the homogeneous coordinates $z_1$, $z_2$ and $z_8$ do not lie on this plane.
The projection along the plane maps $z_2$ to $\zeta_1$ and $z_1\,z_8$ to $\zeta_2$,
where $\zeta_1$ and $\zeta_2$ are the homogeneous coordinates of the $\mbb P^1$.
This projection does indeed define a toric homomorphism because SR$_{\mbb P^1}=\{\zeta_1\,\zeta_2\}\simeq\{z_1\,z_2,\,z_2\,z_8\}$
is a subset of~\eqref{eq:model2nddP8:sr-ideal}. Over the point $\zeta_1=z_2=0$,
we get a generic K3 fibre and at $\zeta_2=0$ the fibre becomes the sum of $D_1$ and $D_8$.

We finally stress that for this fibration we also have global sections of the base.
Given that the two $\mbb P^1$'s turn out to be $\mc C_{2\,\mbb{P}^1}=D_4\cap D_6$, the volume of the base is:
\be
\textmd{Vol}({\rm base})= \tfrac 12 \,\textmd{Vol}(\mc C_{2\,\mbb{P}^1})
=\tfrac 12\,\int_{ D_3\cap D_6} \imath^* J=\tfrac 12\,D_4\cdot D_6\cdot J=r_2 =t_2-2\,t_4.
\ee
Let us summarise the results of our divisor analysis:
\be
 \begin{array}{|c|c|c|c|}
\hline
\textmd{divisor classes} & D_1 & D_2\simeq D_1 + D_8 \\ \hline
{\rm variety} & \textmd{dP}_8 & \textmd{K3} \\ \hline
{\rm volume} & \tfrac 92 r_1^2 & 4\,(r_1 + r_4)\,(2\,r_1 + r_3 + 2\,r_4) \\ \hline
\end{array}\,.\nonumber
\ee

\subsubsection{Orientifolding}

We briefly discuss the  orientifold with
$z_7\leftrightarrow -z_7$ involution. The fixed point set of the ambient space is given by:
\bea
\{\textmd{Fixed}\}\big|_{z_7\leftrightarrow -z_7}
&=&\{D_7,\,D_1\cdot D_5\cdot D_8,\,D_1\cdot D_5\cdot D_6\cdot D_8\}\,.
\eea
However, only $D_7$ and $D_1\cdot D_5\cdot D_8$ intersect the invariant hypersurface,
and so we get an $O7$-plane and an $O3$-plane.

Regarding the Calabi-Yau hypersurface, also the K3 fibre and the del Pezzo eight surface
become less generic in the orientifold case even if, together with all the other toric divisors,
they do not degenerate.

The stringy Euler characteristic of the geometry of the F-theory uplift is 14136. This matches the $D3$-tadpole of a single Whitney brane configuration including the $O3$-plane.

\section{List of all the K3 fibrations with diagonal del Pezzo divisors}
\label{sec:the-list}

In this appendix we give the full list of the 158 polytopes
that admit a K3-fibred Calabi-Yau three-fold with $h^{1,1}=4$ K\"ahler moduli
and at least one `diagonal' del Pezzo divisor.

The lattice polytopes are described in terms of their weight matrices.
All their topological data can be obtained via the C program \emph{mori.x} \cite{Braun:2011ik}.
For example, the command to obtain the information about the Calabi-Yau three-fold
coming from the triangulation of the polytope number 94 is:

\begin{verbatim}
echo '8 4 1 0 1 1 1 0  5 2 1 0 0 0 1 1  12 6 2 1 0 1 2 0' |./mori.x -a
\end{verbatim}

Notice that the example of section~\ref{sec:theexemplar} corresponds to the polytope number 1,
the example discussed in section~\ref{sec:6.1} is the 94th polytope,
the example of section~\ref{sec:6.2} corresponds to the polytope number 136,
the example described in section~\ref{sec:A.1} is given by the 147th polytope,
while the example of section~\ref{sec:A.2} corresponds to the polytope number 79.
We stress that the difference between the weight matrices of the polytopes listed here
and those defining the toric ambient varieties of the examples presented in the main text,
is due to the fact that we are dealing with reflexive lattice polytopes which
can be defined without having to specify all the weights.

\bigskip

\begin{center}
\begin{tabular}{c||c|c|c}
  \# & $\Sigma_i w_i$ & $w_i$ & $\mbb{Z}_2$ action \\
  \hline\hline
  1 & 8 & 2 2 2 1 1 & 0 0 1 1 0 \\
  2 & 8 & 2 2 2 1 1 & 1 1 1 1 0  \\
  3 & 12 & 2 2 6 1 1 & 1 0 0 1 0 \\
  4 & 12 & 2 2 6 1 1 & 1 1 1 1 0 \\
  5 & 12 & 2 2 6 1 1 & 0 0 1 1 0
\end{tabular}

\bigskip

\bigskip

\begin{tabular}{c||c|c|c|c}
  \# & $\Sigma_i w^A_i$ & $w^A_i$ & $\Sigma_i w^B_i$ & $w^B_i$ \\
  \hline\hline
  6 & 8 & 2 2 1 1 2 0 & 2  & 0 0 1 0 0 1  \\
  7 & 6 & 1 3 1 0 0 1 & 4 & 0 1 0 1 2 0 \\
  8 & 8 & 1 3 1 1 0 2 & 6 & 1 3 0 1 1 0  \\
9 & 6 & 2 1 2 0 0 1 & 6 & 3 1 0 1 1 0  \\
  10 & 6 & 1 2 1 1 0 1 & 12 & 3 4 1 0 2 2 \\
  11 & 8 & 4 1 1 0 1 1 & 18 & 9 3 0 1 3 2 \\
  12 & 4 & 2 0 1 0 0 1 & 12 & 6 2 1 1 2 0 \\
  13 & 5 & 1 1 1 0 1 1 & 12 & 3 3 2 1 3 0 \\
14 & 6 & 1 1 1 0 1 2 & 12 & 3 1 3 2 3 0 \\
15 & 6 & 1 0 3 1 1 0 & 12 & 2 1 5 2 0 2 \\
  16 & 12 & 3 3 3 1 0 2 & 12 & 3 3 0 2 3 1 \\
  17 & 14 & 1 7 3 0 1 2 & 16 & 1 8 4 1 2 0 \\
18 & 10 & 5 1 0 2 1 1 & 18 & 9 3 2 0 1 3  \\
19 & 12 & 6 2 0 2 1 1 & 14 & 7 2 1 2 0 2 \\
20 & 18 & 3 9 3 1 0 2 & 18 & 3 9 0 2 3 1 \\
21 & 18 & 6 9 1 0 1 1 & 24 & 8 12 0 1 2 1
\end{tabular}

\begin{tabular}{c||c|c|c|c|c}
  \# & $\Sigma_i w^A_i$ & $w^A_i$ & $\Sigma_i w^B_i$ & $w^B_i$ & $\mbb{Z}_2$ action on B \\
  \hline\hline
  22&4 &1 1 1 0 0 1 & 2 &0 0 0 1 1 0 & 1 1 1 1 0 0 \\
23 & 4 &1 1 1 0 0 1 & 4 &1 1 0 1 1 0 & 0 0 1 1 0 0 \\
24 &4 &1 1 1 0 0 1 & 4 & 1 1 0 1 1 0 & 1 0 0 1 0 0  \\
  25 & 6 & 1 3 1 0 1 0 & 4 &  0 2 0 1 0 1 & 0 1 0 1 0 0 \\
26 & 6 & 1 3 0 1 1 0 & 4 & 0 2 1 0 0 1 & 0 0 1 1 0 0 \\
27&6 &1 3 1 0 0 1 & 6 &1 3 0 1 1 0 & 0 0 1 1 0 0 \\
28 &6 &1 3 1 0 0 1 & 6 &1 3 0 1 1 0 & 1 0 0 1 0 0 \\
29 & 8 & 4 1 0 1 1 1 & 6 & 3 1 1 0 0 1 & 1 1 1 1 0 0  \\
30 &8 &4 1 1 1 0 1 & 6 &3 1 0 1 1 0 & 1 0 1 0 0 0
\end{tabular}

\bigskip

\bigskip

\begin{tabular}{c||c|c|c|c|c|c}
  \# & $\Sigma_i w^A_i$ & $w^A_i$ & $\Sigma_i w^B_i$ & $w^B_i$ & $\Sigma_i w^C_i$ & $w^C_i$ \\
  \hline\hline
  31 &12 &1 1 6 2 0 0 2 & 6& 1 0 3 0 1 0 1 & 6& 0 1 3 1 0 1 0 \\
32&10& 1 5 1 0 1 0 2 & 6& 0 3 0 1 1 0 1 & 8& 2 4 1 0 0 1 0 \\
33&5 &1 1 1 0 1 0 1 & 3& 0 1 1 0 0 1 0 & 8& 1 2 2 1 2 0 0 \\
34 &6& 2 1 1 1 0 1 0 & 3& 1 0 1 0 0 0 1 & 6 &2 2 0 0 1 1 0 \\
35\sphantom 6 &1 2 1 0 0 1 1 & 6 &2 1 0 1 0 1 1 & 6& 2 0 0 2 1 0 1 \\
36\sphantom 8& 2 1 1 2 0 0 2 & 4 &1 1 0 0 1 0 1 & 4& 1 0 1 1 0 1 0 \\
37\sphantom 8 &1 2 2 0 2 1 0 & 4 &1 1 1 0 0 0 1 & 4 &1 0 0 2 0 1 0 \\
38\sphantom 5& 1 1 1 0 1 0 1&  4 &1 1 0 1 0 1 0 & 6& 1 1 1 0 0 1 2 \\
39\sphantom 6 &1 1 0 2 1 1 0 & 3 &0 0 0 1 1 0 1 & 4& 1 1 1 0 0 1 0 \\
40\sphantom 7& 2 1 1 0 2 0 1  &6 &2 1 0 1 2 0 0 & 8& 2 1 2 1 0 2 0 \\
41\sphantom 10& 1 5 1 0 0 1 2&  4 &0 2 1 0 1 0 0&  8& 2 4 0 1 0 1 0 \\
42\sphantom 5 &1 1 1 0 0 1 1&  2 &0 0 0 0 1 0 1 & 5& 1 1 0 1 1 1 0 \\
43\sphantom 8& 2 2 1 0 0 1 2 & 2 &0 0 0 0 1 0 1 & 4& 1 1 0 1 0 1 0 \\
44\sphantom 5 &1 1 1 0 1 0 1  &4 &1 1 1 0 0 1 0  &8& 2 1 2 1 2 0 0 \\
45\sphantom 6& 1 1 0 1 2 1 0&  2 &0 0 0 0 1 0 1 & 4 &1 1 1 0 0 1 0 \\
46\sphantom 7 &1 3 1 1 0 1 0 & 6 &1 3 0 1 1 0 0  &7& 1 3 1 0 1 0 1 \\
47\sphantom 10& 1 5 1 0 1 0 2 & 6 &1 3 1 0 0 1 0  &6& 0 3 0 1 1 0 1 \\
48\sphantom 8& 1 4 1 0 1 0 1 & 6 &1 3 0 1 0 1 0 & 10 &1 5 1 0 0 1 2 \\
49\sphantom 6& 1 2 1 0 1 0 1&  2 &0 1 0 0 0 1 0  &10 &1 4 2 1 2 0 0 \\
50\sphantom 8 &1 4 1 1 0 0 1 & 10 &1 5 2 0 0 1 1  &12& 2 6 1 2 1 0 0 \\
51\sphantom 8 &1 3 0 1 1 2 0  &6 &1 3 1 0 1 0 0 & 10 &1 5 0 1 1 0 2 \\
52\sphantom 6 &1 2 1 0 0 2 0  &6 &1 3 0 1 1 0 0 & 8 &1 4 1 0 0 0 2 \\
53\sphantom 8 &1 4 0 1 1 0 1  &6 &1 3 1 0 0 1 0  &14& 2 7 2 0 1 0 2 \\
54\sphantom 12 &6 1 2 0 1 0 2 & 6& 3 1 0 1 0 0 1 & 8& 4 1 0 0 1 2 0 \\
55\sphantom 4 &1 1 1 0 0 0 1 & 4 &1 1 0 1 0 1 0 & 4& 2 0 0 0 1 1 0 \\
56\sphantom 7 &2 2 1 0 1 1 0 & 3 &1 1 0 0 0 0 1 & 8& 2 2 0 1 2 1 0 \\
57\sphantom 14 &7 3 1 0 0 1 2&  8 &4 2 1 0 1 0 0&  8& 4 2 0 1 0 1 0 \\
58\sphantom 8 &4 1 0 1 1 0 1  &8 &4 2 1 0 0 1 0 & 10& 5 1 1 0 1 0 2 \\
59\sphantom 7 &2 2 1 0 0 1 1&  4 &1 1 1 0 1 0 0 & 6 &2 2 0 1 0 1 0 \\
60\sphantom 8 &4 1 1 0 1 0 1 & 8 &4 1 0 1 2 0 0 & 10& 5 1 2 0 1 1 0 \\
\end{tabular}

\begin{tabular}{c||c|c|c|c|c|c}
  \# & $\Sigma_i w^A_i$ & $w^A_i$ & $\Sigma_i w^B_i$ & $w^B_i$ & $\Sigma_i w^C_i$ & $w^C_i$ \\
  \hline\hline
  61\sphantom 10& 2 5 0 1 1 0 1 & 8 &2 4 1 0 0 1 0 & 14& 3 7 1 0 1 0 2 \\
62\sphantom 8 &1 4 1 1 0 1 0 & 8 &2 4 0 0 1 1 0 & 10 &0 5 2 1 0 1 1 \\
    63\sphantom 10 &5 1 0 1 2 1 0&  4 &2 0 0 0 1 0 1&  8 &4 2 1 0 0 1 0 \\
64\sphantom 6 &1 2 0 1 1 1 0 & 5 &1 2 1 0 0 1 0 & 6 &1 3 1 0 0 0 1 \\
65\sphantom 5 &2 1 0 1 0 0 1 & 4 &2 0 0 0 1 1 0 & 5 &2 0 1 0 1 0 1 \\
66\sphantom 10 &2 5 0 1 0 2 0&  4 &0 2 1 0 0 0 1 & 10& 0 5 0 1 2 0 2 \\
67\sphantom 7 &3 0 1 1 1 1 0& 4 &2 1 0 0 0 0 1 & 5 &2 0 0 1 1 0 1 \\
67\sphantom 5 &1 2 0 1 1 0 0 & 3 &0 1 1 0 0 1 0 & 6 &1 3 0 0 1 0 1 \\
67\sphantom 8 &4 1 0 1 1 1 0  &6 &3 1 1 0 0 1 0 & 6 &3 0 1 0 1 0 1 \\
70\sphantom 14 &7 2 0 2 2 1 0  &6 &3 1 0 0 1 0 1 & 6& 3 0 2 0 0 1 0 \\
71\sphantom 6 &1 1 2 1 1 0 0&  6 &1 1 3 0 0 1 0 & 8 &1 1 4 0 1 0 1 \\
72\sphantom 6 &1 1 2 1 0 1 0 & 6 &1 0 3 1 1 0 0 & 7 &1 0 3 1 0 1 1 \\
73\sphantom 14& 2 7 1 0 0 2 2&  6& 1 3 0 1 0 1 0&  10& 2 5 1 0 2 0 0 \\
74\sphantom 12 &2 2 5 2 0 1 0 & 6 &1 0 3 1 1 0 0 & 14 &2 0 7 2 0 1 2 \\
75\sphantom 6 &1 3 1 0 1 0 0  &4 &0 1 0 1 0 2 0 & 6 &0 3 0 1 0 0 2 \\
76\sphantom 7 &3 1 1 0 1 0 1 & 3 &1 0 1 0 0 1 0 & 6 &3 1 0 1 1 0 0 \\
77\sphantom 6 &1 3 1 0 1 0 0 & 2 &0 0 0 1 0 1 0 & 4 &0 2 0 1 0 0 1 \\
78\sphantom 6 &1 3 1 0 1 0 0 & 2 &0 0 0 1 0 1 0 & 3 &0 1 0 0 0 1 1 \\
79\sphantom 12 &6 2 1 0 1 2 0 & 4 &2 1 0 0 0 0 1 & 6 &3 0 1 1 0 1 0 \\
80\sphantom 8 &4 1 1 0 0 1 1 & 4 &2 0 0 0 1 0 1 & 8 &4 1 0 1 1 1 0 \\
81\sphantom 5 &1 2 1 0 0 1 0 & 5 &1 2 0 1 1 0 0 & 6 &1 3 0 0 1 0 1 \\
82\sphantom 6 &3 0 1 0 1 1 0 & 3 &1 1 0 0 0 0 1 & 4 &2 0 0 1 1 0 0 \\
83\sphantom 7 &1 3 1 0 1 1 0 & 5 &1 2 0 1 1 0 0 & 6 &1 3 0 0 1 0 1 \\
84\sphantom 8 &4 1 1 0 0 1 1 & 6 &3 1 0 1 0 1 0 & 6& 3 0 2 0 1 0 0 \\
85\sphantom 7 &1 3 0 1 1 1 0 & 8 &1 4 0 1 1 0 1 & 12& 2 6 1 1 2 0 0 \\
86\sphantom 10& 5 2 0 1 0 1 1 & 8 &4 2 1 0 0 1 0 & 14& 7 3 0 2 1 1 0 \\
87\sphantom 10 &2 5 0 1 1 1 0 & 6 &1 3 0 0 1 0 1&  10 &2 5 1 0 0 2 0 \\
88\sphantom 14 &2 7 2 0 0 2 1 & 6 &1 3 1 0 1 0 0 & 6 &1 3 0 1 0 1 0 \\
89\sphantom 8 &4 1 1 0 0 1 1 & 6 &3 1 0 1 0 1 0 & 8 &4 2 1 0 1 0 0 \\
90\sphantom 10 &5 2 0 1 1 1 0 & 4 &2 0 0 0 1 0 1 & 8 &4 2 1 0 0 1 0 \\
91\sphantom 14 &7 3 0 1 2 1 0 & 6 &3 1 0 0 1 0 1  &8 &4 2 1 0 0 1 0 \\
92\sphantom 7 &3 1 1 0 1 0 1 & 4 &2 0 0 1 1 0 0&  6 &3 0 1 0 1 1 0 \\
93\sphantom 8 &1 4 0 0 1 1 1 & 10 &2 5 1 0 0 2 0&  12& 2 6 0 1 1 2 0 \\
94\sphantom 8 &4 1 0 1 1 1 0 & 5 &2 1 0 0 0 1 1  &12& 6 2 1 0 1 2 0 \\
95\sphantom 8 &1 4 1 0 1 0 1 & 4 &0 2 1 0 0 1 0&  12& 1 6 2 1 2 0 0 \\
96\sphantom 5 &2 0 0 1 1 0 1 & 4 &2 1 0 0 0 0 1 & 6 &3 0 1 0 0 1 1 \\
97\sphantom 11& 2 5 1 0 2 1 0 & 5 &1 2 0 1 1 0 0 & 6& 1 3 0 0 1 0 1 \\
98\sphantom 11 &2 5 2 1 1 0 0 & 5 &1 2 1 0 0 1 0  &12& 2 6 2 1 0 0 1 \\
99\sphantom 10 &5 2 1 0 1 1 0 & 4 &2 1 0 0 0 0 1 & 12 &6 2 0 1 2 1 0 \\
100\sphantom 10 &2 5 1 0 0 1 1 & 6 &1 3 1 0 1 0 0  &8 &2 4 0 1 0 1 0 \\
101\sphantom 12 &2 6 0 0 1 2 1 & 10 &2 5 1 0 0 2 0  &16& 3 8 0 1 1 3 0 \\
102\sphantom 18 &6 9 0 1 1 1 0 & 12 &4 6 1 0 0 1 0&  12 &4 6 0 0 1 0 1 \\
103\sphantom 8 &4 1 1 0 1 0 1 & 6 &3 1 1 0 0 1 0 & 12& 6 2 1 1 2 0 0 \\
104\sphantom 6 &3 1 0 0 1 0 1 & 6& 3 0 1 1 0 0 1  &8& 4 0 0 2 0 1 1 \\
\end{tabular}

\begin{tabular}{c||c|c|c|c|c|c}
  \# & $\Sigma_i w^A_i$ & $w^A_i$ & $\Sigma_i w^B_i$ & $w^B_i$ & $\Sigma_i w^C_i$ & $w^C_i$ \\
  \hline\hline
  105\sphantom 10& 5 1 0 1 2 1 0 & 4& 2 0 0 0 1 0 1 & 6& 3 1 1 0 0 1 0 \\
106\sphantom 8 &1 1 4 1 0 1 0  &4 &0 1 2 0 0 0 1 & 8 &2 0 4 0 1 1 0 \\
107\sphantom 8 &1 4 0 1 1 1 0 & 4& 0 2 0 0 1 0 1&  10& 2 5 1 0 0 2 0
\end{tabular}

\bigskip

\bigskip

\begin{tabular}{c||c|c|c|c|c|c|c|c}
  \# & $\Sigma_A$ & $w^A_i$ & $\Sigma_B$ & $w^B_i$ & $\Sigma_C$ & $w^C_i$ & $\Sigma_D$ & $w^D_i$ \\
  \hline\hline
108\sphantom 4 & 1 1 0 0 0 1 0 1 & 4& 1 0 0 1 1 0 0 1  &4& 1 0 0 1 0 1 1 0  &4& 0 1 1 0 1 0 0 1 \\
109\sphantom 5& 1 1 0 0 1 1 0 1  &4& 1 1 0 1 0 1 0 0  &4& 1 1 0 0 1 0 1 0  &4& 1 0 1 1 0 0 1 0 \\
110\sphantom 10& 5 1 0 1 0 1 0 2  &4& 2 1 0 0 0 0 1 0  &4& 2 0 1 0 0 1 0 0 & 4& 2 0 0 1 1 0 0 0 \\
111\sphantom 3& 1 0 0 1 0 0 0 1  &3& 1 0 0 0 1 0 1 0  &3& 0 1 1 0 0 0 1 0  &3& 0 1 0 1 0 1 0 0 \\
112\sphantom 5& 1 1 0 1 0 1 1 0  &2& 0 1 0 0 0 0 0 1  &3& 0 0 1 0 0 0 1 1  &4& 1 1 0 0 1 1 0 0 \\
113\sphantom 5& 1 1 1 1 0 1 0 0  &2& 0 1 0 0 0 0 1 0  &3& 1 0 0 0 0 0 1 1  &5& 0 2 1 0 1 1 0 0 \\
114\sphantom 5& 1 1 1 0 1 0 0 1  &2& 0 0 0 0 0 0 1 1  &4& 1 1 1 0 0 1 0 0  &4& 1 1 0 1 1 0 0 0 \\
115\sphantom 4& 1 1 0 1 0 0 1 0  &2& 0 0 1 0 0 0 0 1  &4& 1 0 1 0 0 1 1 0  &4& 0 1 1 0 1 0 1 0 \\
116\sphantom 5&1 1 1 0 0 0 1 1  &3& 1 0 0 0 1 1 0 0  &4& 1 1 0 1 0 0 1 0  &4 &1 1 0 0 0 1 0 1 \\
117\sphantom 5 &1 1 1 0 1 0 1 0  &2& 0 0 0 1 1 0 0 0  &3& 1 1 0 0 0 1 0 0  &4& 1 1 0 0 0 0 1 1 \\
118\sphantom 4 &1 1 0 0 1 0 0 1  &3& 1 1 0 0 0 0 1 0  &3& 1 0 1 0 0 1 0 0  &4 &1 0 0 1 1 1 0 0 \\
119\sphantom 5& 1 1 0 1 1 0 1 0  &2& 0 0 0 1 0 0 0 1  &4& 1 1 1 0 1 0 0 0  &5& 0 1 1 1 1 1 0 0 \\
120\sphantom 5 &1 1 1 0 0 1 0 1  &4& 1 1 1 0 0 0 1 0  &4& 1 1 0 0 1 1 0 0  &4 &1 0 1 1 0 1 0 0 \\
121\sphantom 6 &1 2 0 0 1 1 1 0  &2& 0 1 0 0 0 0 0 1  &2& 0 0 0 1 0 1 0 0  &4& 1 1 1 0 0 0 1 0 \\
122\sphantom 6 &2 1 0 1 0 1 1 0  &3& 1 1 0 0 0 0 0 1  &3& 1 0 1 0 0 0 1 0  &3& 1 0 0 0 1 1 0 0 \\
123\sphantom 5 &1 0 1 1 0 1 1 0  &3& 1 0 0 1 1 0 0 0  &4 &1 1 0 1 0 0 1 0  &4& 1 0 0 1 0 1 0 1 \\
124\sphantom 10 &5 1 0 1 0 1 2 0  &4& 2 1 0 0 0 0 0 1  &4 &2 0 1 0 0 1 0 0  &6& 3 0 0 1 1 0 1 0 \\
125\sphantom 7 &2 2 0 1 1 1 0 0  &3& 1 1 0 0 0 0 1 0  &4 &1 1 1 0 0 1 0 0  &4 &1 1 0 0 1 0 0 1 \\
126\sphantom 6 &2 1 0 1 1 1 0 0 & 2& 1 0 0 0 0 0 0 1  &2 &0 0 1 0 0 1 0 0 & 5& 2 1 0 0 1 0 1 0 \\
127\sphantom 8 &1 0 4 0 1 1 0 1  &6& 1 0 3 1 0 1 0 0  &6 &1 0 3 0 1 0 1 0  &6& 0 1 3 1 0 0 1 0 \\
128\sphantom 6 &2 1 1 0 0 1 1 0  &2& 1 0 0 0 0 0 0 1  &5 &2 1 0 0 1 1 0 0 & 5& 2 0 1 1 0 1 0 0 \\
129\sphantom 7 &1 3 1 0 0 1 1 0  &6& 1 3 0 0 1 0 1 0  &7 &1 3 0 1 1 1 0 0 & 8& 1 4 1 0 0 0 1 1 \\
130\sphantom 10 &1 5 1 0 2 1 0 0  &4& 1 2 0 0 0 0 0 1  &4 &0 2 0 1 1 0 0 0  &4 &0 2 0 0 0 1 1 0 \\
131\sphantom 7& 3 1 0 1 1 1 0 0  &2& 1 0 0 0 0 0 0 1 & 3& 1 0 1 0 0 1 0 0  &5& 2 1 0 0 1 0 1 0 \\
132\sphantom 8& 4 1 0 0 1 1 1 0 & 4 &2 1 0 0 0 0 0 1  &4 &2 0 0 1 1 0 0 0  &10& 5 1 1 0 1 2 0 0 \\
133\sphantom 10& 5 0 1 1 0 2 1 0&  4 &2 1 0 0 0 0 1 0 & 4 &2 0 0 1 1 0 0 0  &6& 3 0 0 1 0 1 0 1 \\
134\sphantom 6 &1 2 1 1 0 1 0 0 & 5 &1 2 0 1 1 0 0 0 & 6 &1 3 0 0 1 0 0 1  &7& 1 3 0 1 0 1 1 0 \\
135\sphantom 5& 1 0 2 1 0 1 0 0 & 4 &1 0 2 0 0 0 1 0 & 5 &0 1 2 1 1 0 0 0&  6& 1 0 3 0 0 1 0 1 \\
136\sphantom 8& 4 1 1 1 0 1 0 0 & 4 &2 1 0 0 0 0 0 1 & 6 &3 1 0 1 1 0 0 0 & 6& 3 0 1 0 0 0 1 1 \\
137\sphantom 5& 1 2 0 1 1 0 0 0 & 3 &0 1 1 0 0 1 0 0 & 4 &0 2 1 0 0 0 0 1  &6& 1 3 0 0 1 0 1 0 \\
138\sphantom 6& 1 2 1 1 1 0 0 0 & 6 &1 3 0 1 0 1 0 0 & 7& 1 3 1 1 0 0 0 1  &7& 1 3 0 1 1 0 1 0 \\
139\sphantom 8& 0 4 1 1 0 1 1 0 & 4 &1 2 0 0 0 0 0 1 & 6 &1 3 0 1 0 0 1 0&  6& 0 3 0 1 1 1 0 0
\end{tabular}

\begin{tabular}{c||c|c|c|c|c|c|c|c}
  \# & $\Sigma_A$ & $w^A_i$ & $\Sigma_B$ & $w^B_i$ & $\Sigma_C$ & $w^C_i$ & $\Sigma_D$ & $w^D_i$  \\
  \hline\hline
  140\sphantom 8& 1 4 1 0 1 0 0 1&  4 &0 2 0 0 0 0 1 1 & 6 &1 3 1 0 0 1 0 0 & 6& 1 3 0 1 1 0 0 0 \\
141\sphantom 6& 3 1 0 0 1 0 1 0 & 4 &2 0 1 0 0 0 0 1 & 6 &3 0 1 1 0 0 1 0  &6& 3 0 1 0 1 1 0 0 \\
142\sphantom 7& 1 3 0 1 1 1 0 0  &3 &0 1 1 0 0 1 0 0 & 4 &0 2 1 0 0 0 0 1  &6& 1 3 0 0 1 0 1 0 \\
143\sphantom 7 &3 1 0 1 1 1 0 0&  5 &2 1 0 0 1 0 1 0 & 6 &3 1 1 0 0 1 0 0  &6& 3 1 0 0 0 0 1 1 \\
144\sphantom 6 &1 3 1 0 1 0 0 0 & 2 &0 0 0 1 0 1 0 0 & 3 &0 1 0 1 0 0 0 1  &3& 0 1 0 0 0 1 1 0 \\
  145\sphantom 8 &4 1 1 0 1 0 0 1  &4 &2 0 1 0 0 0 1 0 & 4 &2 0 0 0 0 1 0 1 & 6 &3 1 0 1 1 0 0 0 \\
146\sphantom 5 &1 2 1 0 0 1 0 0 & 5 &1 2 0 1 1 0 0 0 & 6 &1 3 1 0 0 0 0 1 & 6& 1 3 0 0 1 0 1 0 \\
147\sphantom 8 &1 4 0 0 1 1 1 0&  4 &1 2 0 0 0 0 0 1 & 4 &0 2 1 0 0 0 1 0 & 4 &0 2 0 1 0 1 0 0 \\
148\sphantom 8 &1 4 1 0 0 1 1 0 & 4 &0 2 1 0 1 0 0 0 & 5 &1 2 0 1 0 1 0 0  &6& 1 3 0 0 0 1 0 1 \\
149\sphantom 6 &3 1 0 0 1 0 0 1  &4 &2 1 0 0 0 0 1 0 & 4 &2 0 1 0 0 1 0 0  &6& 3 0 0 1 1 1 0 0 \\
150\sphantom 8 &4 1 1 0 0 1 0 1 & 6 &3 1 1 0 0 0 1 0 & 6 &3 1 0 0 1 1 0 0 & 6& 3 0 1 1 0 1 0 0 \\
151\sphantom 8 &4 1 1 0 1 0 1 0 & 4 &2 0 1 0 0 0 0 1 & 6 &3 1 0 1 1 0 0 0  &8& 4 0 1 1 1 1 0 0 \\
152\sphantom 10 &5 2 0 1 0 1 1 0 & 4& 2 1 0 0 0 0 0 1&  4& 2 0 0 0 1 1 0 0  &6& 3 1 1 0 0 0 1 0 \\
153\sphantom 7 &1 3 1 0 1 1 0 0 & 5 &1 2 0 1 1 0 0 0  &6 &1 3 0 0 1 0 1 0  &8& 1 4 1 0 1 0 0 1 \\
154\sphantom 7 &1 3 0 1 1 1 0 0 & 6 &1 3 1 0 0 1 0 0  &6 &1 3 0 0 1 0 1 0  &8& 1 4 0 0 1 1 0 1 \\
155\sphantom 8 &1 4 0 1 1 0 1 0 & 3 &0 1 0 1 0 1 0 0  &4 &0 2 0 1 0 0 0 1  &6& 1 3 1 0 1 0 0 0 \\
156\sphantom 8 &4 1 0 1 0 1 1 0 & 4 &2 0 1 0 0 1 0 0  &4 &2 0 0 1 1 0 0 0  &6& 3 0 0 1 0 0 1 1 \\
157\sphantom 8 &4 0 1 1 0 1 1 0 & 4& 2 1 0 0 0 0 1 0  &6 &3 0 1 0 0 0 1 1  &6& 3 0 0 1 1 0 1 0 \\
158\sphantom 10 &5 2 0 1 1 1 0 0 & 4& 2 1 0 0 0 0 1 0  &6 &3 1 1 0 0 1 0 0  &6& 3 1 0 0 1 0 0 1
\end{tabular}
\end{center}

\providecommand{\href}[2]{#2}\begingroup\raggedright\endgroup
%

\end{document}